\documentclass[aps,prd,twocolumn,superscriptaddress]{revtex4-2}
\usepackage{graphicx}
\usepackage{amsmath}
\usepackage{amssymb}
\usepackage{hyperref}
\usepackage{xcolor}
\usepackage{array}
\usepackage{booktabs}

\usepackage{float}
\usepackage{tikz}
\usepackage{comment}
\usepackage{xcolor}
\usepackage[most]{tcolorbox}
\definecolor{traj1}{rgb}{0.25,0.41,0.88}
\definecolor{traj2}{rgb}{0.18,0.55,0.34}
\definecolor{traj3}{rgb}{1.00,0.39,0.28}
\definecolor{traj4}{rgb}{0.73,0.33,0.83}
\definecolor{traj5}{rgb}{0.72,0.53,0.04}
\definecolor{traj6}{rgb}{0.00,0.66,0.80}

\begin{document}

\title{Baryon and lepton asymmetries across the cosmic QCD transition: effects of a 3D first-order surface}

\author{Lorenzo Formaggio}\thanks{Corresponding author: lformagg@cougarnet.uh.edu}
\affiliation{Department of Physics, University of Houston, Houston, TX 77204, USA }

\author{Francesco Di Clemente}
\affiliation{Department of Physics, University of Houston, Houston, TX 77204, USA }
\author{Tristan Gyure}
\affiliation{Department of Physics, University of Houston, Houston, TX 77204, USA }

\author{Hitansh Shah}
\affiliation{Department of Physics, University of Houston, Houston, TX 77204, USA }
\affiliation{Institute for Theoretical Physics, Goethe University,
Max-von-Laue-Strasse 1, 60438 Frankfurt am Main, Germany}

\author{Roman Poberezhniuk}
\affiliation{Department of Physics, University of Houston, Houston, TX 77204, USA }

\author{Alessandro Drago}
\affiliation{INFN, Sezione di Ferrara, via Saragat 1, I-44122 Ferrara, Italy}
\affiliation{Department of Physics and Earth Science, University of Ferrara, via Saragat 1, I-44122 Ferrara, Italy}

\author{Volodymyr Vovchenko}
\affiliation{Department of Physics, University of Houston, Houston, TX 77204, USA }
 
\author{Claudia Ratti}
\affiliation{Department of Physics, University of Houston, Houston, TX 77204, USA }
\date{\today}
\begin{abstract}
We investigate the cosmological QCD phase transition by calculating the cosmic trajectories from the five equations that impose the conservation of the baryon and lepton asymmetries, the charge neutrality of the Universe, and $\beta$ equilibrium.
We perform the calculation using for the first time the four-dimensional $(T, \mu_B, \mu_Q, \mu_S)$ Constant Entropy Contour equation of state, which for $\mu_S = 0$ ($\beta$-equilibrium) features a critical line and a first-order phase transition (FOPT) surface in the three-dimensional $(T, \mu_B, \mu_Q)$ space.
We first show that, within the standard cosmological scenario with baryon asymmetry $b_{\text{obs}}=8.65\cdot 10^{-11}$, the lepton asymmetries alone can never induce a FOPT.
We then explore the non-standard cosmological scenario with large values of $b$ and find that reaching a FOPT requires $b\gtrsim0.028$ during the QCD epoch.
Such a value is compatible with observations only if the baryon asymmetry is diluted to $b_{\text{obs}}$ after the transition, which a period of little inflation would provide with $N\approx6.5$ $e$-folds of expansion.
Finally, we study the phenomenological imprints on the equation of state and primordial black hole (PBH) formation, showing that pion condensation at large lepton asymmetries drives $c_s^2 > 1/3$ and shifts the PBH peak to sub-solar masses ($0.3-0.8\,M_\odot$), while the large-$b$ FOPT yields a narrow spike around $4.2\,M_\odot$.
\end{abstract}

\maketitle

\section{Introduction}
Cosmologically, the QCD phase transition has long been assumed to occur at vanishing chemical potential ($\mu_B \approx 0$) due to the small observed baryon asymmetry, $b = (n_B - n_{\bar{B}})/s \approx 8.65 \times 10^{-11}$ \cite{Planck:2018vyg}. 
At small baryochemical potential, lattice QCD calculations show the transition to be a smooth crossover, which produces virtually no cosmological relics~\cite{Aoki_2006,Arslandok:2023utm,Borsanyi:2025dyp}. 
However, this standard picture has been challenged recently, by the detection of a stochastic gravitational wave background (SGWB) in the nanohertz frequency range reported by Pulsar Timing Array (PTA) collaborations \cite{Afzal_2023}. 
While a smooth crossover cannot generate an observable gravitational wave signal, a first-order phase transition (FOPT) may provide a compelling cosmological explanation for this background via bubble collisions, sound waves, and magnetohydrodynamic turbulence in the cosmic plasma. 
This observational breakthrough opens the real possibility that the cosmological QCD phase transition was not a crossover, but a first-order transition, leading to the question: can realistic cosmological mechanisms induce a QCD FOPT in the early Universe?

The answer to this question depends on the cosmic trajectory on the QCD phase diagram, which is not determined by the baryon asymmetry alone. 
As first shown in Ref. \cite{Schwarz:2009ii}, the trajectory of the Universe is shaped not only by baryon asymmetry, but also by the three individual lepton asymmetries $\ell_e, \ell_\mu, \ell_\tau$. 
Through the condition of global electric charge neutrality, any net charge carried by the leptons must be balanced by the hadronic sector, which induces non-zero values for both the electric charge chemical potential $\mu_Q$ and the baryon chemical potential $\mu_B$. 
Subsequent investigations using lattice QCD based equations of state (EoS) Taylor-expanded up to second order \cite{Oldengott_2017,wygas2018cosmic,PhysRevD.105.123533} and fourth order \cite{Formaggio:2025nde,diclemente2025upperboundcosmicbaryon} demonstrated that, within observationally allowed lepton asymmetry bounds ($|\ell| \le 10^{-2}$ and $|\ell_e| \le 10^{-3}$), $\mu_B$ cannot reach high enough values, keeping the transition in the crossover regime. 

Adopting a functional QCD framework, Ref. \cite{Gao_2022} calculated the cosmic trajectories using an EoS where the critical end point (CEP) is located at $T_{\mathrm{CEP}} = 125\,\mathrm{MeV}$ and $\mu_{B,\mathrm{CEP}} = 333\,\mathrm{MeV}$ ($\mu_u = \mu_d = 200\,\mathrm{MeV}$. In this framework, they demonstrated that a FOPT can be induced in the early Universe given sufficiently large flavor lepton asymmetries (e.g., $|\ell_\mu| = |\ell_\tau| \gtrsim 7\times 10^{-2}$ for $\ell_e = 0$), which drive the electric charge neutrality condition to generate substantial quark chemical potentials and steer the cosmic evolution across the first-order transition line.

While providing an important proof of principle, existing treatments in the literature have two major limitations. 
First, by construction, Taylor-expanded lattice EoSs do not feature a critical point or a first-order phase transition line. 
Second, the exploration of the electric charge chemical potential direction remains incomplete, a limitation that affects both lattice calculations and the functional QCD framework of Ref. \cite{Gao_2022}. 
In fact, neither approach fully captures the effects introduced by an evolution in the finite-$\mu_Q$ direction. On the lattice side, the analysis of the QCD phase transition was mostly limited to the $\mu_B-T$ plane, neglecting the effects of a non-zero electric charge chemical potential, with Taylor expansions breaking down when large lepton asymmetries push $|\mu_Q|$ to hundreds of $\mathrm{MeV}$. In Ref. \cite{Gao_2022}, on the other hand, the condition for entering the FOPT regime was evaluated through a simplified threshold on individual flavor chemical potentials ($|\mu_u|=|\mu_d|\ge 111\,\mathrm{MeV}$, rather than through a fully multidimensional equation of state $P(T, \mu_B, \mu_Q)$ that incorporates the shift of the critical line induced by $\mu_Q$. 

To provide a conclusive answer, it is necessary to compute cosmic trajectories within a self-consistent thermodynamic framework that predicts and incorporates a critical line and a first-order phase transition surface across the complete multidimensional $(T, \mu_B, \mu_Q)$ space. In this work, we achieve this by incorporating the Constant Entropy Contour (CEC) EoS~\cite{Gyure:2026stu}. 
Based on lattice QCD data at $\mu = 0$, the CEC EoS approach mitigates the limitations of standard Taylor series by extrapolating the constant entropy density contours on the QCD phase diagram, reconstructing the entropy from them, and then obtaining a complete thermodynamic surface that incorporates critical scaling around the CEP in the full $(T, \mu_B, \mu_Q)$ domain (setting $\mu_S = 0$ because of $\beta$-equilibrium).

Using this 3D framework, we demonstrate that within the standard cosmological scenario, lepton asymmetries alone are never sufficient to drive the cosmic trajectory into the first-order phase transition: the critical line shifts to higher $\mu_B$ at finite $\mu_Q$ in such a way that the trajectory always remains inside the crossover region.

Instead, entering the FOPT requires relaxing the constraint on the primordial baryon asymmetry to a minimal value of $b \gtrsim 0.028$ at the QCD scale. To restore this value to the observed $b_{\mathrm{obs}} \approx 8.65 \times 10^{-11}$, the excess baryon asymmetry must be diluted by a factor of $D \approx 3.2 \times 10^8$. If this entropy injection were driven by a period of little inflation \cite{Boeckel_2012} during the transition, $N \approx 6.5$ $e$-folds of supercooled expansion would be required. Within the present EoS, however, the QCD transition itself cannot provide this dilution, because its metastable phase persists only a few MeV below coexistence and therefore cannot support significant supercooling that instead would be needed for this mechanism to work.

This paper is structured as follows. In Sec.~\ref{cosmology} we formulate the system of conservation equations, $\beta$-equilibrium, and describe the cosmological mechanism for the baryo-lepto genesis, the different scenarios for the asymmetries and  the little inflation mechanism. In Sec.~\ref{EoS} we present the CEC EoS and its critical structure in the $(T, \mu_B, \mu_Q)$ space. In Sec.~\ref{results} we analyze the resulting cosmic trajectories, identifying the conditions for crossover, FOPT, and pion condensation. Finally, in Sec.~\ref{conclusions} we summarize our conclusions and provide an outlook.

\section{Cosmology}
\label{cosmology}
The cosmic trajectories on the QCD phase diagram (the space of temperature, baryon and electric charge chemical potentials) represent the evolution of the early Universe along the strong interaction domination phase. At any moment of its evolution, the Universe is a well-defined configuration of chemical potentials and temperature (a point in the phase diagram), and as the time passes, its temperature decreases and such evolution becomes a line (trajectory) on the QCD phase diagram. These trajectories are calculated by solving a system of conservation equations that maps the evolution of the chemical potentials as a function of the temperature $T$~\cite{Formaggio:2025nde,wygas2018cosmic,PhysRevD.105.123533,Gao_2022}. In practice, the temperature acts as the independent evolution parameter: for each discrete thermal step, the system is solved numerically. Since we are interested in the QCD phase transition, the temperature regime considered in this work spans from $80\, \mathrm{MeV}$ up to $250 \,\mathrm{MeV}$. The solution depends on the chosen EoS, which defines the relationships between the chemical potentials and the temperature, and on the parameters of the system: the three lepton asymmetries and, when its constraint is relaxed, the baryon asymmetry. Indeed, depending on the EoS and the set of parameters, the trajectory can either pass through a first-order phase transition or a crossover. The high sensitivity of the system to these model-dependent choices necessitates a comprehensive overview of the underlying equations, the physical constraints governing the parameters, and the selection of the EoS.
In the following subsection, we outline the mathematical definition of the system and discuss why $\beta$-equilibrium conditions must be considered.

\subsection{Cosmic Trajectories: How to calculate them}
\subsubsection{Cosmic Trajectory Equations}
\label{cosmic trajectories equation}
At vanishing chemical potentials, the QCD crossover occurs at $T \simeq 156\, \text{MeV}$. To fully understand the transition across all regimes, we compute the corresponding cosmic trajectories in the range $80\,\text{MeV} \lesssim T \lesssim 250\,\text{MeV}$.

In this regime, the baryon asymmetry, $b = (n_B(T,\mu_B,\mu_Q)-n_{\bar B}(T,\mu_B,\mu_Q))/s$, can be assumed to be constant, fixed at the baryogenesis. In the same way, the Electro-Weak (EW) phase transition occurs at $T_{EW}\approx 160 \,
\mathrm{GeV}$, fixing the three lepton asymmetries $\ell_i$, where $i=e,\mu,\tau$. The neutrino oscillations, responsible for the variation of the three lepton asymmetries, become effective at temperature lower than $10\,\mathrm{MeV}$, meaning that $\ell_i$ are kept constant over the QCD phase transition. Finally, we can always assume the electric charge neutrality of the Universe. Therefore, we have the following five conservation equations:

\begin{equation}
    b=\sum_i B_i \frac{\bar{n}_i }{s}
    \label{bcons}
\end{equation}
\begin{equation}
     0=q =\sum_i Q_i \frac{\bar{n}_i}{s} 
     \label{qneutr}
\end{equation}
\begin{equation}
\ell_\alpha=\frac{\bar{n}_\alpha+\bar{n}_{\nu_\alpha}}{s} \quad \alpha={e,\mu,\tau} \label{lepasym}
\end{equation}
where $i$ refers to all the particles participating in the system (leptons, photons and gluons have just $B_i=0$, $b=8.65\cdot 10^{-11}$ is the baryon asymmetry, $B_i$ is the baryon charge of each particle, $Q_i$ is the electric charge of each particle and $s$ is the total entropy density of the Universe. Solving this set of equations at different values of temperature we obtain the cosmic trajectory. Since we have five conservation equations, only five chemical potentials are independent. However, the thermodynamic system contains many more particle species, each characterized by its own chemical potential. The relevant particles at these temperatures are the charged leptons and their neutrinos, photons, gluons, and the four lightest quark flavors. The charm quark has been shown to provide a relevant contribution only for temperatures above $T \sim 180\,\mathrm{MeV}$ \cite{Formaggio:2025nde}. While our considered temperature range reaches up to $T \simeq 250\, \mathrm{MeV}$, the effects of the charm quark in this narrow upper band are small enough that they do not qualitatively alter the cosmological evolution. Thus, we neglect its contribution in this study. In this work, a $2+1$ flavor EoS is adopted. The remaining particles, namely the electroweak gauge bosons, the Higgs boson, and the heavier quarks, are too massive to contribute appreciably to the thermodynamics of the Universe at these temperatures.

\subsubsection{Beta Equilibrium}
The QCD phase epoch lasts $\sim 1 \,\mathrm{\mu s }$ while the beta process time scale is $\sim 10^{-2}\mathrm{\mu s}$, therefore they thermalize the system. Such condition reduces the number of independent variables to exactly five as follows:
\begin{align}
 \mu_B & =\mu_{u}+2\mu_{d}\\
  \mu_Q& =\mu_{u}-\mu_{d}\\
  \mu_S& =0\\ 
\mu_\mu& =\mu_e+\mu_{\nu_\mu}-\mu_{\nu_e} \\
\mu_\tau& =\mu_e+\mu_{\nu_\tau}-\mu_{\nu_e} \\
\mu_{d}& =\mu_{u}+\mu_e-\mu_{\nu_e} \\
\mu_{d}& =\mu_{s }
\end{align}
where the first two lines are the definition of the basis.
Note that $\mu_S$ is the strangeness chemical potential, and not the chemical potential of the strange quark, which instead is $\mu_{s}=\frac{1}{3}\mu_B-\frac{1}{3}\mu_Q-\mu_S=\mu_{d}$.  Now we have to choose a basis for our five chemical potentials we want to work with, and since the equations we solve are the baryon asymmetry conservation, charge neutrality and the three lepton asymmetry conservation, the most natural basis to choose is the following:
\begin{equation}
      \{ \mu_B,\mu_Q, \mu_{\nu e}, \mu_{\nu \mu}, \mu_{\nu \tau}\}
\end{equation}
Note that if instead of $\mu_B$ and $\mu_Q$ we would have used $\mu_{up}$ and $\mu_{down}$, the solutions would not change at all.

As shown in previous studies~\cite{Formaggio:2025nde,wygas2018cosmic,PhysRevD.105.123533,Gao_2022}, the behavior of the solution to the system \ref{bcons}--\ref{lepasym} depends sensitively on the system parameters, particularly on the three lepton asymmetries. The goal of this paper is to show not only how the trajectories behave when a $3D$ critical line is present in the QCD EoS, but also how the baryon asymmetry $b$ could induce a cosmological FOPT. For this reason, we wish to emphasize the importance of the mechanisms that lead to a non-zero baryon asymmetry and lepton asymmetries in the early Universe, as well as the possible mechanisms that can modify their values.

\subsection{Cosmological Set-Up}
\label{Cosmological set-Up}
The observed baryon asymmetry is well constrained, $b=8.65\cdot 10^{-11}$ \cite{Planck:2018vyg}, while we have only upper bounds for the lepton asymmetries: $|\ell_e|\leq 10^{-3}$ and $|l|=|\ell_e+\ell_\mu+\ell_\tau|\leq10^{-2}$~\cite{Oldengott_2017}. As already discussed in Ref.~\cite{Formaggio:2025nde}, these constraints arise from Big Bang Nucleosynthesis (BBN), which is sensitive to the electron-neutrino abundance in the Universe, and from observations of the Cosmic Microwave Background (CMB), which are sensitive to the expansion rate of the Universe. The latter depends on the total energy density, which is indeed sensitive to the total lepton asymmetry. This is because a non-zero lepton asymmetry introduces neutrino chemical potentials, which increase the total neutrino energy density via Fermi-Dirac statistics. This additional energy density affects the cosmic expansion rate and leaves a measurable imprint on the CMB spectrum.

Following this, we revisit the main mechanisms of baryo-leptogenesis to explore how the relaxation of the baryon asymmetry constraint allows for a broader scan of the parameter space.

\subsubsection{Baryo-Lepto genesis}
\label{Baryo-lepto genesis}
In the early Universe, particles are produced during the reheating phase subsequent to inflation. This process releases a massive amount of entropy. Since cosmological asymmetries are the ratio between the net densities and the entropy density, this entropy injection, combined with the exponential expansion of the Universe,  dilutes any pre-existing primordial asymmetry to zero, effectively resetting both the baryon and lepton asymmetries: $b=l=0$ \cite{Kolb:1990vq,Dine:2003}. Then a mechanism that generates the asymmetries is required. The idea is well described in the framework of the baryo-lepto genesis, described as follows.

At the quantum level, the global baryon number $B$ and lepton number $L$ are violated by electroweak anomalies \cite{tHooft:1976rip}. 
This phenomenon is unsuppressed at high temperatures in the Early Universe. 
Crucially, since the anomalies affect both quantum numbers equally ($\Delta B=\Delta L$), the orthogonal combination, namely the difference $B-L$ remains a conserved global charge, while $B+L$ is strongly violated \cite{Weinberg1979, Wilczek1979}. What is generated is actually not the baryon asymmetry, but the lepton asymmetry instead. Indeed, as presented in \cite{Fukugita:1986hr}, much before the EW phase transition, at $T\sim10^9-10^{13}\, \mathrm{GeV}$, the decay of sterile Majorana neutrinos violates the lepton number, generating a non zero lepton asymmetry (leptogenesis). 
Once $\ell$ is generated, the sphaleron processes are responsible to share such asymmetry with the baryon asymmetry $b$ \cite{Kuzmin:1985mm,Kolb:1990vq}. 
At temperatures above the electroweak scale ($T \gtrsim 100\,\mathrm{GeV}$) \cite{Okun:1982ap}, sphaleron processes are unsuppressed. 
Driven by thermal fluctuations, the cosmic plasma frequently transitions between topologically distinct vacuum sectors. 
Because these processes violate baryon $B$ and lepton $L$ numbers while strictly conserving $B-L$, each vacuum configuration corresponds to different values of $B$ and $L$. 
Assuming these transitions are efficient enough to maintain thermal equilibrium in the plasma, a linear relationship between the baryon and lepton numbers is established. 
In terms of baryon asymmetry $b$ and total lepton asymmetry $\ell$ such relationship reads as: $\ell=-\frac{51}{28}b$ \cite{PhysRevD.42.3344}. 
This relationship defines the \textit{Minimal Standard Scenario}, for which the baryon asymmetry is the observed one and the lepton asymmetry is defined through such equation.

This entire mechanism (which highlights why the process is referred to as baryo-leptogenesis) concludes once the electroweak phase transition completes. 
As the Higgs field acquires a non-zero vacuum expectation value, the sphaleron processes become exponentially suppressed. 
Once their interaction rate drops below the Hubble expansion rate ($\Gamma_{\mathrm{sph}} < H$), they fall out of thermal equilibrium and freeze out. 
At this point, no further processes can redistribute the asymmetries between $b$ and $\ell$; their values remain fixed, permanently frozen in the linear relationship derived above \cite{Kajantie_1996}.

Therefore, both the baryon asymmetry and lepton asymmetry are fixed at the end of the EW phase transition. 
Let us now consider the possible mechanism that could produce a variation of $b$ and $\ell$ at lower temperature. 
Regarding the lepton asymmetries, as we already discussed in \autoref{cosmic trajectories equation}, they are affected by the neutrino oscillations, which become effective for $T\lesssim 10\, \mathrm{MeV}$. 
There are other mechanisms that could alter these values, such as the late decay of sterile neutrinos with masses below the EW scale \cite{Eijima_2022}, or medium-induced active-sterile neutrino oscillations, where thermal plasma effects drive chaotic or resonant conversions that dynamically amplify or redistribute the asymmetry at lower temperatures \cite{PhysRevD.53.R5349,Shi_1999}. Another possible mechanism that could produce a huge lepton asymmetry after the EWPT is through the late decay of \textit{L-balls}. However, this mechanism only becomes effective at temperatures lower than the QCD phase transition regime \cite{Kawasaki_2022, PhysRevD.66.043516}, making it irrelevant to our discussion.

Very similar mechanisms could induce the increase of the baryon asymmetry after the EWPT, such as the late decay of \textit{B-balls} \cite{ENQVIST1998309}, WIMP baryogenesis \cite{Cui:2011ab}, or asymmetry-sharing oscillations in dark sectors \cite{Kaplan:2009ag}.
 
Finally, we want to discuss the mechanism of \textit{Little Inflation} \cite{Boeckel_2012}, which could be responsible for the strong dilution of both baryon and lepton asymmetries. 
The idea is that primordial baryo-leptogenesis generates a very high $b$, which can induce a first-order QCD phase transition. 
The transition would first enter an under-cooled regime of the QGP, trapping the system in the false vacuum. 
The constant energy density of this false vacuum drives a short period of accelerated expansion (\textit{Little Inflation}). Once the transition probability is high enough, the system transits into the true vacuum phase (the hadronic gas). 
This transition releases a large amount of latent heat, injecting a huge amount of entropy into the system. 
Such mechanism induces the dilution of both baryon and lepton asymmetries, since they are both normalized by the entropy density, which dramatically increases after the transition. 
The existence of such a long-lived metastable phase within QCD has recently been questioned~\cite{Jung:2026jlz}.

We stress that this dilution acts identically on all asymmetries: a dilution factor $\mathcal{D}$ rescales $b$ and $\ell_i$ by the same amount, leaving the ratios $\ell/b$ and $\ell_i/l$ invariant. It follows that entropy injection alone cannot decorrelate the lepton asymmetry from the baryon one, so that reaching the QCD epoch with a large $b$ and a small $\ell$ requires one of the post-EWPT mechanisms listed above and not just a large $\mathcal{D}$. Second, the bounds on the lepton asymmetries are post-dilution constraints, since both BBN and the CMB probe the Universe after the entropy release. When imposed at the QCD epoch, they must be rescaled by the same factor, $|\ell_{\mathrm{QCD}}|\le 10^{-2}\,b_{\mathrm{QCD}}/b_{\mathrm{obs}}$, and are therefore no longer restrictive once $b_{\mathrm{QCD}}\gg b_{\mathrm{obs}}$.

This mechanism is crucial for the discussion in this paper, as it allows us to investigate the behavior of cosmic trajectories with a very high baryon asymmetry that is subsequently diluted to the observed value after the QCD phase transition. Indeed, it is fundamental to clarify the exact epochs at which $b$ is produced, diluted, and observed. To address this, the next section will discuss this timeline, defining which values for the asymmetries can be considered standard and which ones cannot.
 
\subsubsection{b and l: minimal standard, standard and not standard scenarios}
As introduced in \autoref{Cosmological set-Up}, the baryon asymmetry is measured to be $b = 8.65 \times 10^{-11}$ \cite{Planck:2018vyg}. The earliest epoch at which we can measure this value is during $\mathrm{BBN}$, at a temperature of $T_{\mathrm{BBN}} \sim 1 \,\mathrm{MeV}$. 
This observational constraint is perfectly compatible with scenarios where the QCD phase transition is reached with a significantly higher baryon asymmetry, which is subsequently diluted to the currently observed value well before the onset of $\mathrm{BBN}$.

In the \textit{Minimal Standard Scenario} ($\ell=-\frac{51}{28}b$), the total lepton asymmetry is related to $b$ as introduced above. Nevertheless for the lepton asymmetries we only possess upper bounds: the total lepton asymmetry is constrained to $|l| \le 10^{-2}$ by CMB observations, while the electron lepton asymmetry is limited to $|\ell_e| \le 10^{-3}$ by BBN data. Consequently, we lack a direct measurement for any of these asymmetries at the time of baryogenesis. This implies we cannot verify the relation $\ell=-\frac{51}{28}b$, which may not be true after the EWPT if other mechanisms become efficient after the EWPT and change their values independently.

Therefore, we define three different possible scenarios, in order of theoretical and observational accuracy: \textit{Minimal Standard Scenario}, \textit{Standard Scenario}, \textit{Non Standard Scenario}.

\textit{Minimal Standard Scenario}: the Sphaleron processes are assumed to maintain the thermal equilibrium in the plasma, introducing the relationship between $b$ and $\ell$ introduced in \autoref{Baryo-lepto genesis}. Including the cosmological observation on $b$, we can define this scenario with:
\begin{align}
    &b=8.65\cdot10^{-11}\label{eq:bstandard}\\
    &\ell_{msc}=-51/28\, b=-15.7\cdot 10^{-11}
    \label{eq:lmsc}
\end{align}
Note that $\ell_{msc}$ is the total lepton asymmetry, which can be obtained with several combinations of the individual lepton asymmetries, as long as the constraints on the lepton asymmetries are respected ($|l| \le 10^{-2}$, and $|\ell_e| \le 10^{-3}$). So the symmetric configuration ($\ell_e=\ell_\mu=\ell_\tau=\ell_{msc}/3$) is part of the minimal standard scenario, as well as the asymmetric configuration where $\ell_\tau=-\ell_\mu$ and $\ell_e=\ell_{msc}$. It has been noted that the differences between calculation with $\ell_e=0$ or $\ell_e=\ell_{msc}$ are irrelevant, which allows us to consider also the e-phobic scenario as a minimal standard scenario.

\textit{Standard Scenario}: as done in the previous work, the Standard  Scenario is characterized by the observed baryon asymmetry $b$, without imposing on the total lepton asymmetry the relation to $b$ defined by the sphaleron processes. So here the only constraints we respect are the observed value of $b$ all together with the upper bounds on $\ell_e$ and on the total lepton asymmetry. 

\textit{Non Standard Scenario}: in this last scenario all the constraint on the baryon and lepton asymmetries are relaxed independently, which requires both a large dilution factor and an efficient post-EWPT mechanism among those listed above. In particular in this manuscript we investigate $8.65\cdot10^{-11}\le b\le1$.

\section{Equation of State}
\label{EoS}
In previous works \cite{Formaggio:2025nde,diclemente2025upperboundcosmicbaryon} we approached the problem with different EoSs. First, we assumed the Free QGP model, followed by a $2+1$ flavor Lattice QCD EoS, based on a fourth order Taylor expansion~\cite{PhysRevC.100.064910,abuali2025new4dlatticeqcd,jahan_2025_15123623}, 

extended to $2+1+1$ flavors by incorporating charm-quark susceptibilities calculated in lattice QCD \cite{BAZAVOV2024138520}. For the low temperature regime, the HRG model has been used. In particular, to calculate the cosmic trajectories for this regime  we used the \texttt{Thermal-FIST} package \cite{VOVCHENKO2019295}. As discussed in the Introduction, the main limitations of these previous studies are that the employed EoS did not incorporate the QCD critical point, and the information on its location was either limited to the $\mu_B-T$ plane, neglecting the effects of a non-zero electric charge chemical potential, or taken from external estimates of the 4D critical surface. In this paper, for the first time, we use an equation of state obtained from the extrapolation of constant entropy contours~\cite{Shah:2026mzu,Gyure:2026stu}
that features a critical surface in the 4D space ($T,\mu_B,\mu_Q,\mu_S$). This allows us to have a full analysis of 
the behavior of cosmic trajectories and the cosmological QCD phase transition.
In the next section, we review the main features of the CEC EoS and demonstrate how it reproduces the results obtained in previous works 
using the fourth-order Taylor expansion Lattice QCD EoS.

\begin{figure}
    \centering
    \includegraphics[width=\linewidth]
        {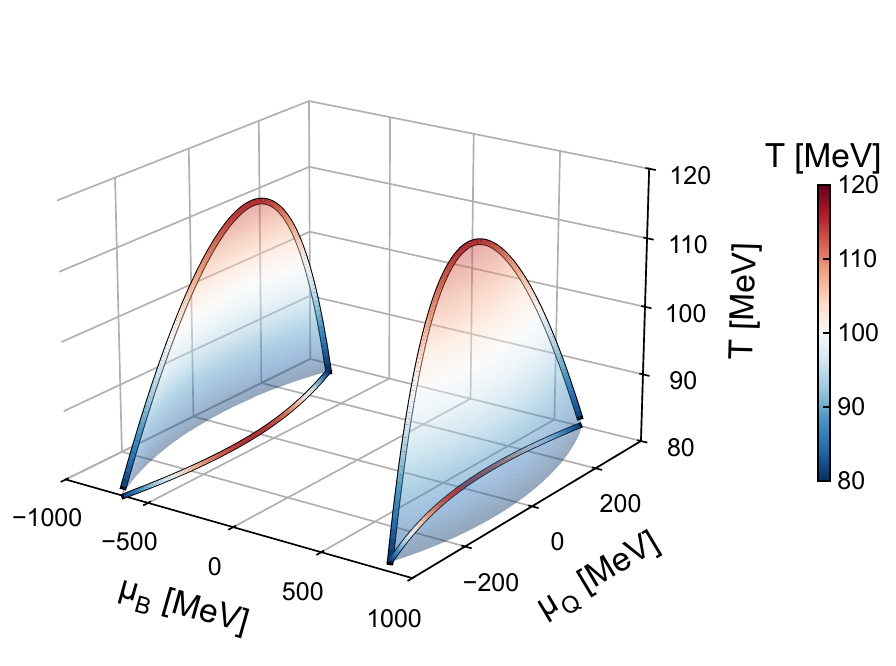}
    \caption{Critical lines and first-order surface in the $(T,\mu_B,\mu_Q)$ space, constructed using the CEC expansion. Critical lines are projected onto the $(\mu_B,\mu_Q)$ plane.}
    \label{fig:first_order_surface}
\end{figure}

\begin{figure*}
    \centering
    \includegraphics[width=\textwidth]
        {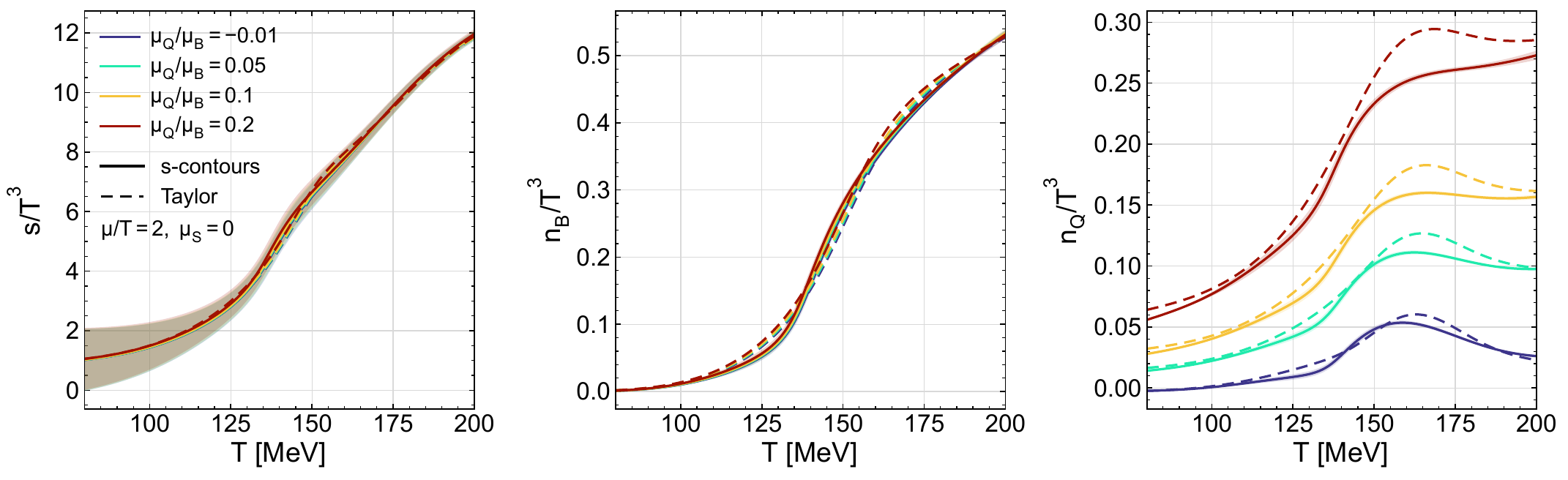}
    \caption{Equation of state from CEC (solid band) compared with Taylor (dashed) expanded normalized entropy density, normalized baryon density, and normalized electric charge density for different values of $\mu_Q/\mu_B$ at $\mu/T=2$ with $\mu_S=0$.
    }
    \label{fig:eos_taylor}
\end{figure*}

\subsection{Constant Entropy Contour EoS}
The equation of state is obtained by the expansion of lattice QCD thermodynamics that follows constant entropy density contours. First introduced in Ref.~\cite{Shah:2024img}, this equation of state yields the entropy density at finite $T$ and $\mu_B$ from:
\begin{equation}
    s(T_s(T_0;\mu_B),\mu_B) = s(T_0,\mu_B=0)
\end{equation}
where $T_s(T_0,\mu_B)$ is the temperature at finite arbitrary $\mu_B$ at fixed value of entropy. This temperature profile expanded into a Taylor series reads:
\begin{equation}
    T_s(\mu_B; T_0) \approx T_0 + \sum_{n=1}^N \alpha_{2n}(T_0) \,\frac{\mu_B^{2n}}{(2\,n)!} + \mathcal{O}\left(\mu_B^{2(N+1)}\right).
    \label{eq:expS}
\end{equation}
Here, the series is truncated at $\mathcal{O}(\mu_B^{2n})$. In Refs.~\cite{Shah:2024img,Shah:2026pue}, the expansion was truncated at $\mathcal{O}(\mu_B^2)$ giving:
\begin{equation}
\label{eq:TsO2}
    T_s(\mu_B; T_0)=T_0+\alpha_2(T_0)\frac{\mu_B^2}{2},
\end{equation}
Here, the expansion coefficient has the form
\begin{align}
    \alpha_2(T_0)
    &=\left(\frac{\partial^2 T}{\partial\mu_B^2}\right)_s\Bigg\vert_{T=T_0,\mu_B=0},
\end{align}
and is calculated using lattice results for second-order baryonic susceptibility and entropy density at vanishing chemical potential. The reason to terminate the expansion at second order for $\mu_B$ was provided in Ref.~\cite{Shah:2026pue,Borsanyi:2025dyp} where it was mentioned that entropy contours at imaginary chemical potential for lattice QCD were approximately quadratic in nature. The critical point in the equation of state can be obtained from conditions involving the derivatives of the expansion coefficient with respect to the temperature. This comes from the fact that the critical point is an inflection point and that the entropy possesses a divergent slope at the CP. When these conditions are applied on the expansion, one gets the following:
\begin{align}
    \mu_{B,c}=\sqrt{-\frac{2}{\alpha_2^\prime(T_{0,c})}} \hspace{0.5cm }, \hspace{0.5cm} \alpha_2^{\prime\prime}(T_{0,c})=0.
\end{align}

However, for our application, we require the QCD equation of state with multiple conserved charges. The expansion of constant entropy-density contours was performed on the full four-dimensional space $(T,\mu_B,\mu_Q,\mu_S)$ in Refs.~\cite{Shah:2026mzu,Gyure:2026stu}. This was obtained by switching the chemical potential space from cartesian to spherical coordinates through the mapping:

\begin{align}
    \mu_B&=\mu\cos\theta,
    \\
    \mu_Q&=\mu\sin\theta\cos\varphi,
    \\
    \mu_S&=\mu\sin\theta\sin\varphi.
\end{align}
Implementing the above equations, reduces the four-dimensional problem to a 2D extrapolation in $(T,\mu)$ that changes with the angular coordinates $(\theta,\varphi)$ spanning the full 4D space. Hence the expansion becomes:
\begin{equation}
\label{eq:TsO24D}
    T_s^{\theta,\varphi}(T_0,\mu)=T_0+\alpha_2^{\theta,\varphi}(T_0)\frac{\mu^2}{2},
\end{equation}
where the coefficient of expansion now depends on the angular coordinates, which basically changes the susceptibility being used based on the direction of expansion. Further details on the expansion coefficient and its components can be found in Refs.~\cite{Shah:2026mzu,Gyure:2026stu}. 

Utilizing the expansion, one obtains the entropy density at finite chemical potentials and temperature. Integrating the entropy density with respect to the temperature provides us with pressure at finite chemical potentials and temperature:
\begin{align}
\label{eq:pressure}
    \int_{T_\text{low}}^T dT^\prime s(T^\prime,\mu_B,\mu_Q,\mu_S)=&P(T,\mu_B,\mu_Q,\mu_S)
    \\&-P(T_\text{low},\mu_B,\mu_Q,\mu_S), \nonumber
\end{align}
where the integration constant $P(T_\text{low},\mu_B,\mu_Q,\mu_S)$ is calculated at $T_\text{low}=80$ MeV using the quantum van der Waals hadron resonance gas~(QvdW-HRG) model \cite{Vovchenko:2016rkn}, augmented with effective mass model to describe pion condensation~\cite{Vovchenko_2021}. The rest of the thermodynamics at finite $\mu$'s can be constructed through the pressure.

Regarding our application for the equation of state to obtain the cosmic trajectories, we set the strangeness chemical potential $\mu_S$ to zero, yielding a critical line in the $(T,\mu_B,\mu_Q)$ space, with a two-dimensional first-order surface, shown in \autoref{fig:first_order_surface}. The CEC equation of state has been shown to agree with already available lattice QCD expansions up to $\mu/T \approx 3.5$ in various directions~\cite{Shah:2026pue,Gyure:2026stu}.
Here, in \autoref{fig:eos_taylor}, we compare this EoS with Taylor expanded lattice QCD thermodynamics at low chemical potential, where we also observe an excellent agreement among the two for the normalized baryon, electric charge, and entropy densities for various ratios of $\mu_Q/\mu_B$.

\section{Cosmological QCD Phase transition}
\label{results}
The main result presented in this manuscript is that a first-order cosmological QCD phase transition cannot be reached without relaxing the constraint on the baryon asymmetry $(b)$, even when the effects of the electric charge chemical potential are taken into account and the EoS presents a critical line in the ($T-\mu_B-\mu_Q$) space. The lepton asymmetries are relevant for the evolution of the cosmological trajectories, but they are never sufficient to drive the system into a FOPT. In contrast, an enhanced baryon asymmetry can lead to a first-order cosmological QCD phase transition. Indeed, when $b\approx0.03$ is assumed, a FOPT is achieved. However, as discussed above, such a scenario requires a mechanism capable of diluting the baryon asymmetry to its observed value after, or possibly during, the transition, such as the little inflation.

In this manuscript we present the cosmic trajectories calculated for different configurations of the parameters:
\begin{itemize}
    \item \textbf{minimal standard scenario:}  $b=8.65\cdot 10^{-11}$, $\ell=\ell_{msc}=-51/28 b$ (the choice for the three individual lepton asymmetries is not relevant in this case);
    \item \textbf{standard scenario:} $b=8.65\cdot 10^{-11}$, symmetric case ($\ell_e=\ell_\mu=\ell_\tau=l/3$), electro-phobic case ($\ell_e=0\, ,\, \ell_\mu=-\ell_\tau$);
    \item \textbf{non standard scenario:} $8.65\cdot10^{-11}\le b\le1$ and $\ell\approx b$.
\end{itemize}
The trajectories for all configurations were calculated using the CEC Equation of State and then directly compared with the corresponding critical line and first-order phase transition surface in the $\mu_B-\mu_Q-T$ space.

\subsection{Standard and minimal standard scenarios}
\begin{figure}[h] 
\centering 
\textbf{Symmetric standard scenario} \\
\vspace{1 mm}
\fbox{%
\footnotesize
\begin{tabular}{ccccc}
\textcolor{traj1}{\rule{0.3cm}{1pt}} $\ell_{msc}$ &
\textcolor{traj2}{\rule{0.3cm}{1pt}} $\ell=10^{-8}$ &
\textcolor{traj3}{\rule{0.3cm}{1pt}} $\ell=10^{-6}$ &
\textcolor{traj4}{\rule{0.3cm}{1pt}} $\ell=10^{-4}$ &
\textcolor{traj5}{\rule{0.3cm}{1pt}} $\ell=10^{-2}$
\end{tabular}%
}

\includegraphics[width=1\columnwidth,trim={0.5cm 2.8cm 0.5cm 4.5cm},clip]{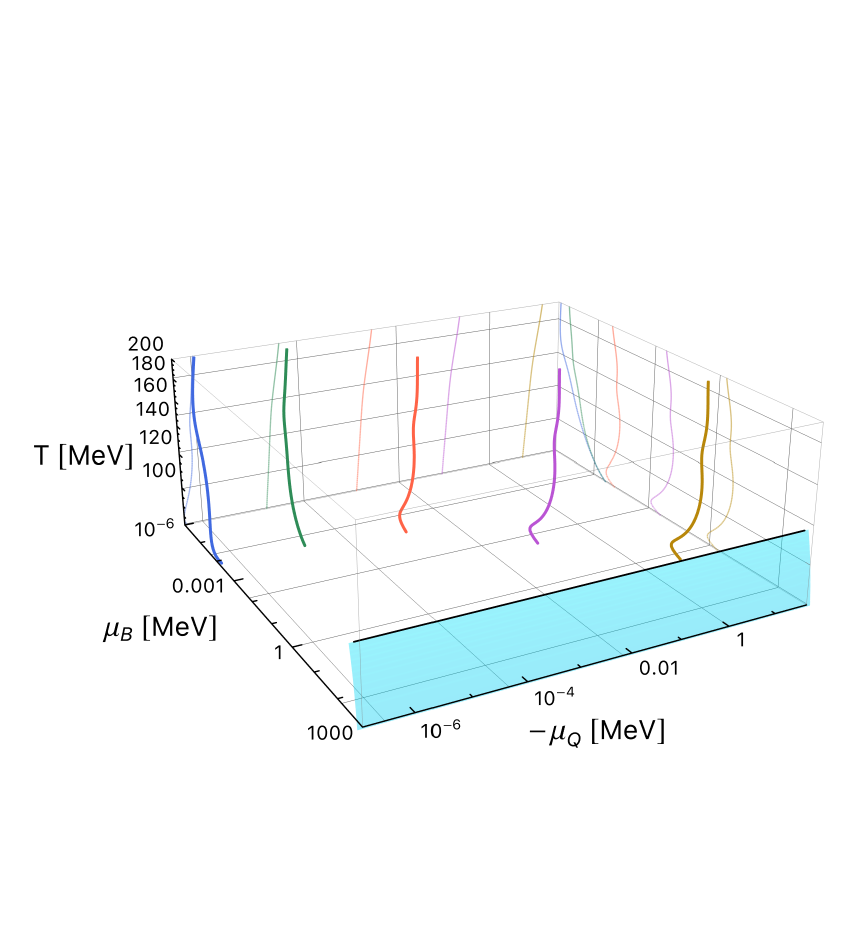} 
\caption{Cosmic trajectories for standard symmetric configuration of the lepton asymmetries ($\ell_e=\ell_\mu=\ell_\tau$) compared to the first-order phase transition. Lighter curves are the orthogonal projections in the $T-\mu_B$ and $T-\mu_Q$ planes} 
\label{standardsym} 
\end{figure}

\begin{figure}[h]
    \centering
    \textbf{Electro-phobic scenario}
    \includegraphics[width=1\columnwidth]{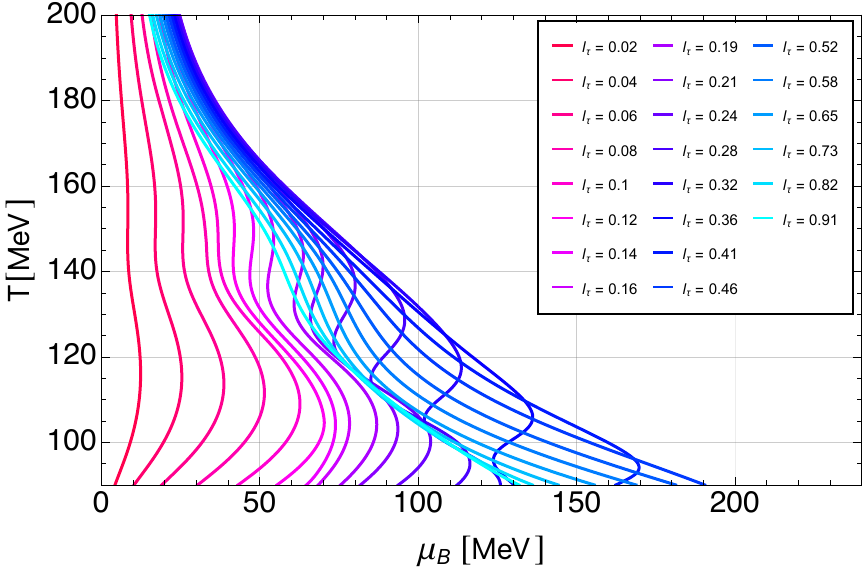}
    \includegraphics[width=1\columnwidth]{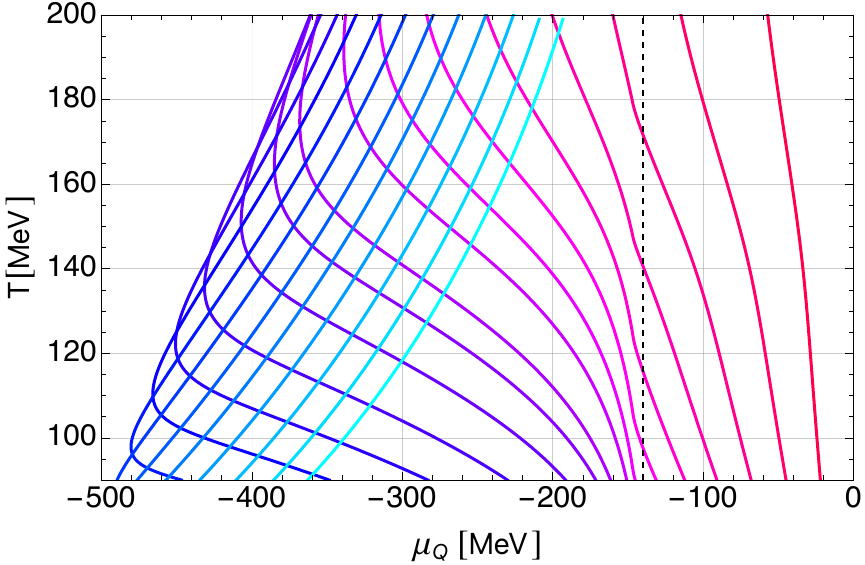}
   \includegraphics[width=1\columnwidth,trim={0.5cm 2.8cm 0.5cm 5.1cm},clip]{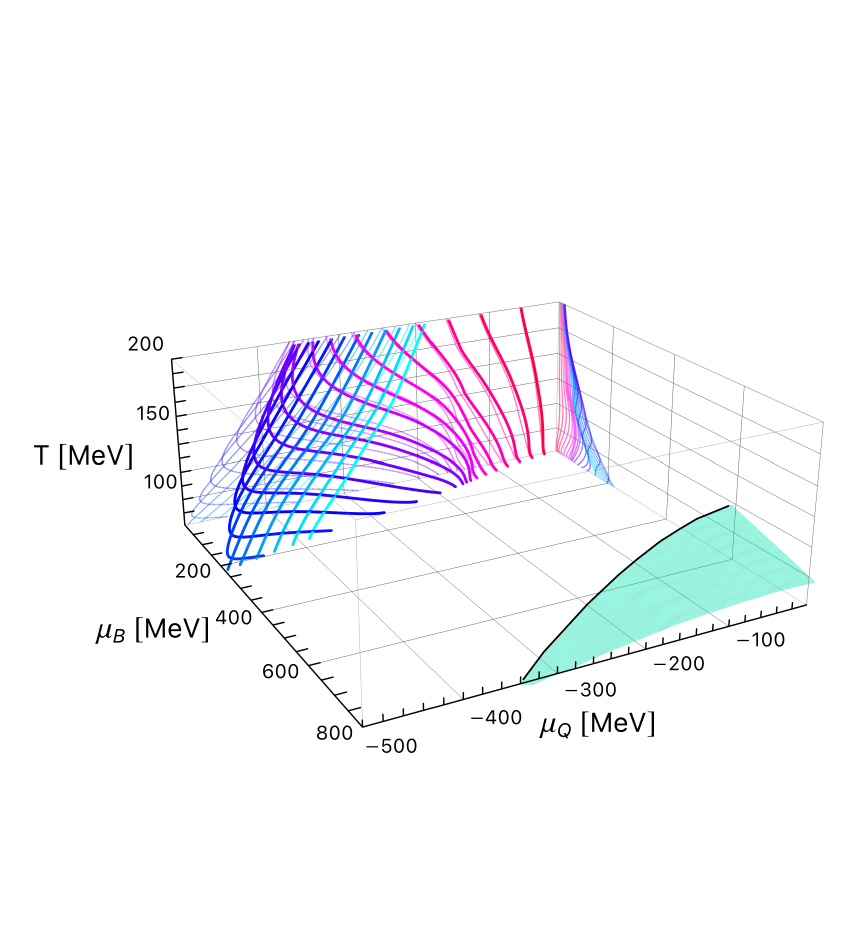}
   \caption{Cosmic trajectories in the $(\mu_B,T)$ and $(-\mu_Q,T)$ planes, together with their three-dimensional representation and the corresponding first-order phase transition surface (light blue), for $\ell_e=0$ and $\ell_\mu=-\ell_\tau$. In the second panel the dashed vertical line is placed at $|\mu_Q|=140\, \mathrm{MeV}\approx m_\pi$ to show the pion condensate formation regime.}
    \label{bounce}
\end{figure}

We first present the results for the trajectories obtained for $b=8.65\cdot10^{-11}$ and the symmetric configuration of the lepton asymmetries, shown in \autoref{standardsym} and \autoref{bounce}, as done in \cite{Formaggio:2025nde}. Note that here the minimal standard scenario is represented with the blue trajectory, obtained with $\ell=\ell_{msc}$. The figure indicates that the critical line and the first-order phase transition surface remain nearly constant as functions of $\mu_Q$, with the electric-charge chemical potential shifting them noticeably along the $\mu_B$ direction only for $|\mu_Q| \gtrsim 100\,\mathrm{MeV}$. Nevertheless, this displacement is insufficient to bring the critical baryon chemical potential within the range reached by the cosmic trajectories. Consequently, the cosmological QCD transition remains a crossover, and no first-order phase transition occurs in this scenario. 

For the e-phobic configuration, we obtain upper bounds on the chemical potentials, exhibiting the same behavior already discussed in \cite{diclemente2025upperboundcosmicbaryon}, as shown in \autoref{bounce}. The figure presents projections onto the planes $\mu_B$--$T$ and $-\mu_Q$--$T$, where the bounce is observed for both chemical potentials. Throughout the temperature range $80\,\mathrm{MeV} \le T \le 250\,\mathrm{MeV}$, the chemical potentials remain bounded by $\mu_B \lesssim 220\,\mathrm{MeV}$ and $|\mu_Q| \lesssim 540\,\mathrm{MeV}$, ensuring that the cosmic trajectories remain always inside the crossover regime. \textit{Note} that this e-phobic scenario is Standard, and not Minimal Standard since the total lepton asymmetry is set to zero. However, to make it minimal standard it would be sufficient to assume $\ell_e=\ell_{msc}$, and so $\ell=\ell_e+\ell_\mu+\ell_\tau=\ell_e=\ell_{msc}$, but this change in $\ell_e$ does not give any relevant difference to the results presented in \autoref{bounce}, as already discussed above. We conclude that \autoref{bounce} presents both minimal standard and standard scenarios.

These results are obtained using an EoS that presents the critical line and first-order surface in 3D, which leads to the conclusion that the lepton asymmetries, whatever the configuration considered, do not induce a cosmological first order QCD phase transition, within the standard and minimal standard cosmological scenarios. This forces us to relax the constraint on the baryon asymmetry, and to investigate the non standard cosmological scenario, as done in the next subsection.

\subsection{Non standard cosmology scenario: how large baryon asymmetry can induce a FOPT}
\label{littleinflation}
The position of the cosmic trajectories in the $\mu_B$--$T$ plane is mainly determined by the QCD sector of the EoS. 
Although this may appear trivial, the leptons play an important role.
Indeed, when we increase the lepton asymmetries in order to drive the trajectory through the first-order phase transition, we are not explicitly increasing the baryon content of the system. 
Rather, we increase the net lepton densities, which in turn enhances the electric charge carried by the leptons. 
Through the charge-neutrality condition, the net baryon density must then increase in order to compensate for this excess charge and preserve overall electric neutrality. As a consequence, the baryon chemical potential is forced to increase as well. By contrast, when we increase $b$, we directly impose a larger net baryon density, and therefore a larger baryon chemical potential.

\begin{figure}[h] 
\centering 
\textbf{Non standard scenario - $\ell=-b$} \\
\vspace{1 mm}
\fbox{%
\footnotesize
\begin{tabular}{ccccc}
\textcolor{traj1}{\rule{0.3cm}{1pt}} $b=10^{-10}$ &
\textcolor{traj2}{\rule{0.3cm}{1pt}} $b=10^{-8}$ &
\textcolor{traj3}{\rule{0.3cm}{1pt}} $b=10^{-6}$ &
\textcolor{traj4}{\rule{0.3cm}{1pt}} $b=10^{-4}$ &
\textcolor{traj5}{\rule{0.3cm}{1pt}} $b=10^{-2}$
\end{tabular}%
}\\
\includegraphics[width=1\columnwidth,trim={0.5cm 2.8cm 0.5cm 4.5cm},clip]{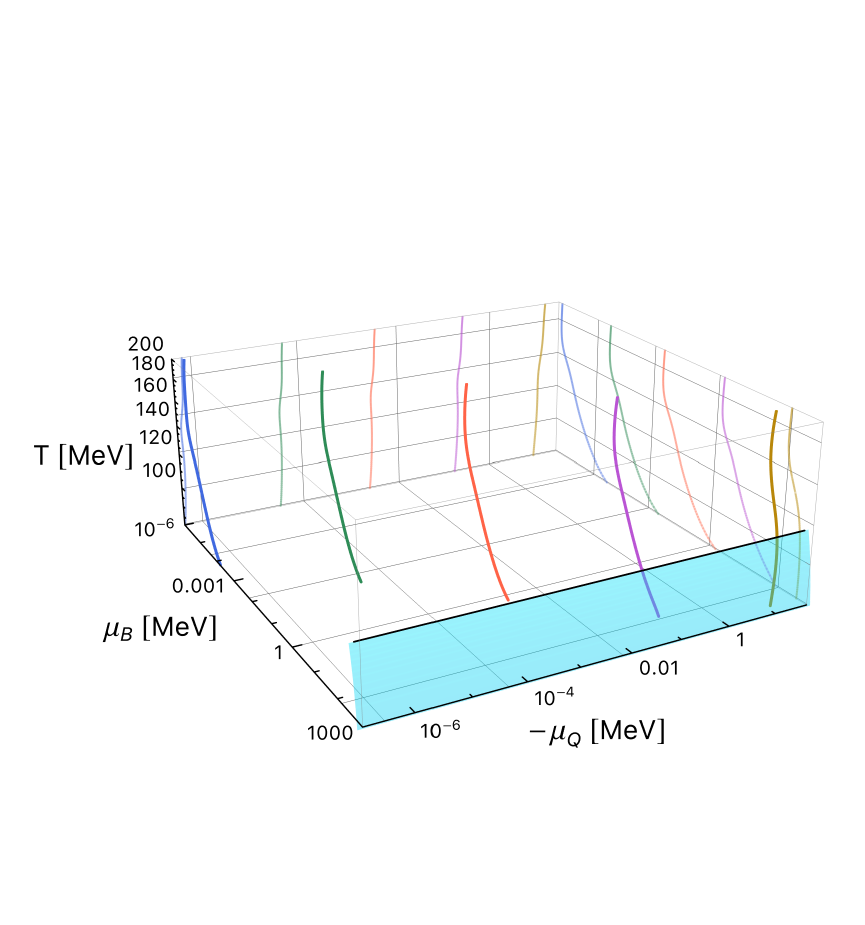} \caption{Cosmic trajectories for different values of the baryon asymmetry. For each value of $b$, $\ell_e=\ell_\mu=\ell_\tau=l/3$ and $b=-l$ is assumed.} \label{wideb} 
\end{figure}

The effects of these mechanisms can be observed in the different configurations considered in this manuscript. In \autoref{bounce} and \autoref{standardsym}, the baryon asymmetry is kept fixed to $b=8.65\cdot 10^{-11}$, while $\ell$ is varied, so that only the charge-balance effect is operative. In this scenario, for a fixed $\mu_B$, the corresponding $|\mu_Q|$ is significantly larger, with
$2 \lesssim \frac{|\mu_Q|}{\mu_B} \lesssim 5$. In contrast, in \autoref{wideb} we increase $b$, assuming for simplicity that $\ell=-b$ (i.e., $\ell_i=-b/3$), although sphaleron processes would instead imply $\ell_{\rm msc}=-\frac{51}{28}b$. In this case, the increase in the net baryon density is direct, and we obtain $\frac{|\mu_Q|}{\mu_B} \sim 0.15$.

\begin{figure}
    \centering
    \includegraphics[width=1\linewidth]{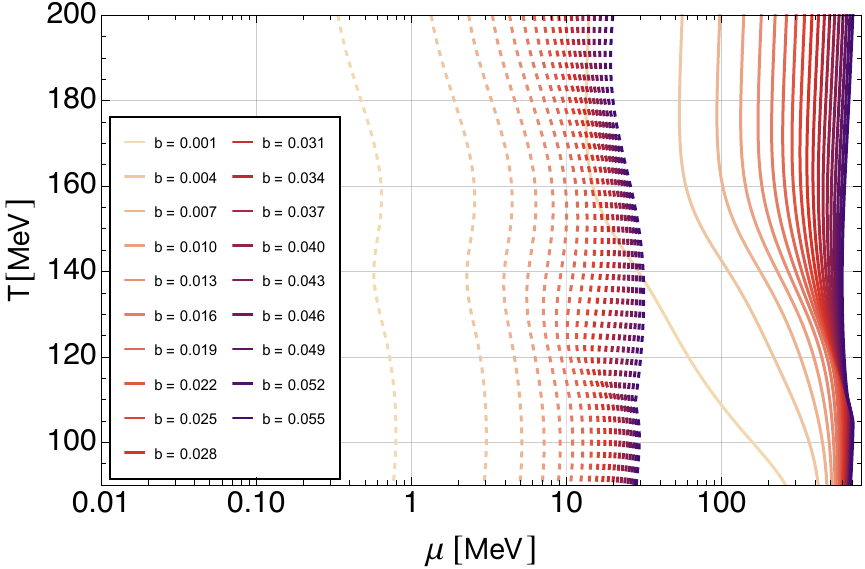}
    \caption{Cosmic trajectories calculated with the CEC EoS for different values of $b$ and constant $\ell=-10\cdot 10^{-10}$. Solid lines: $\mu_B$; Dot-dashed lines: $-\mu_Q$}
    \label{fig:bvaryngl}
\end{figure}

Finally, if only $b$ is increased while keeping $\ell$ of the same order of magnitude as the \textit{minimal standard scenario}, $\mu_Q$ remains even smaller, since baryons provide only a tiny contribution to the electric charge, while the net lepton density is negligible. Indeed, as can be seen from \autoref{fig:bvaryngl}, the ratio becomes
$\frac{|\mu_Q|}{\mu_B} \sim 0.013$,
which is about one order of magnitude smaller than in the case $b\approx -\ell$.

\begin{figure}[h]
    \centering
    \textbf{Non standard scenario: FOPT}
    \includegraphics[width=1\columnwidth]{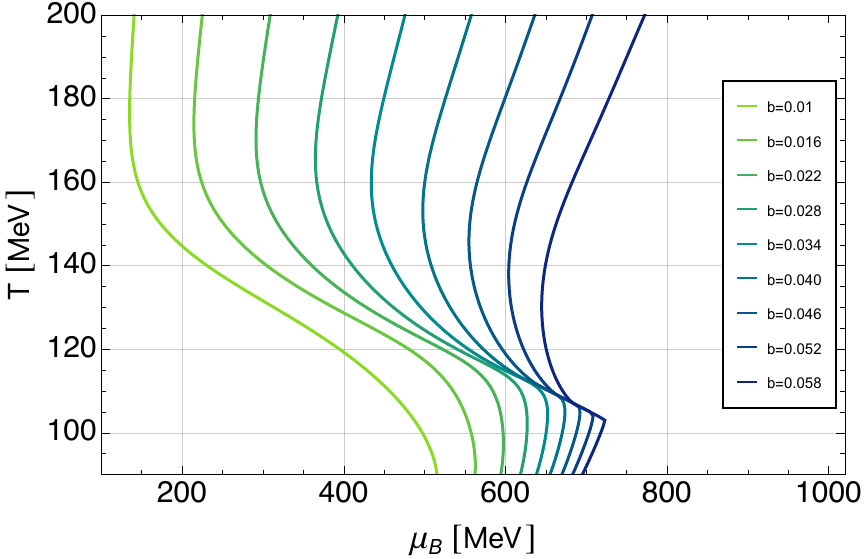}
    \includegraphics[width=1\columnwidth]{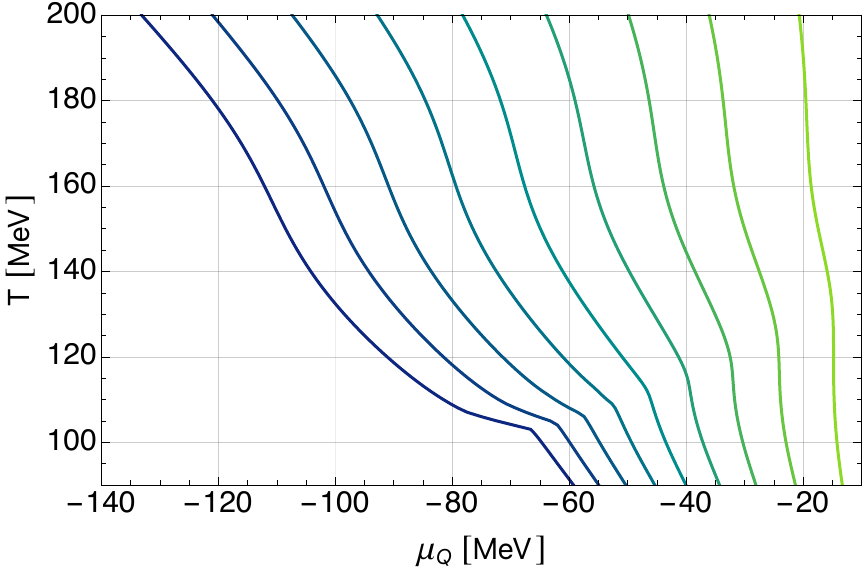}
    \includegraphics[width=1\columnwidth,trim={0.5cm 2.8cm 0.5cm 4.5cm},clip]{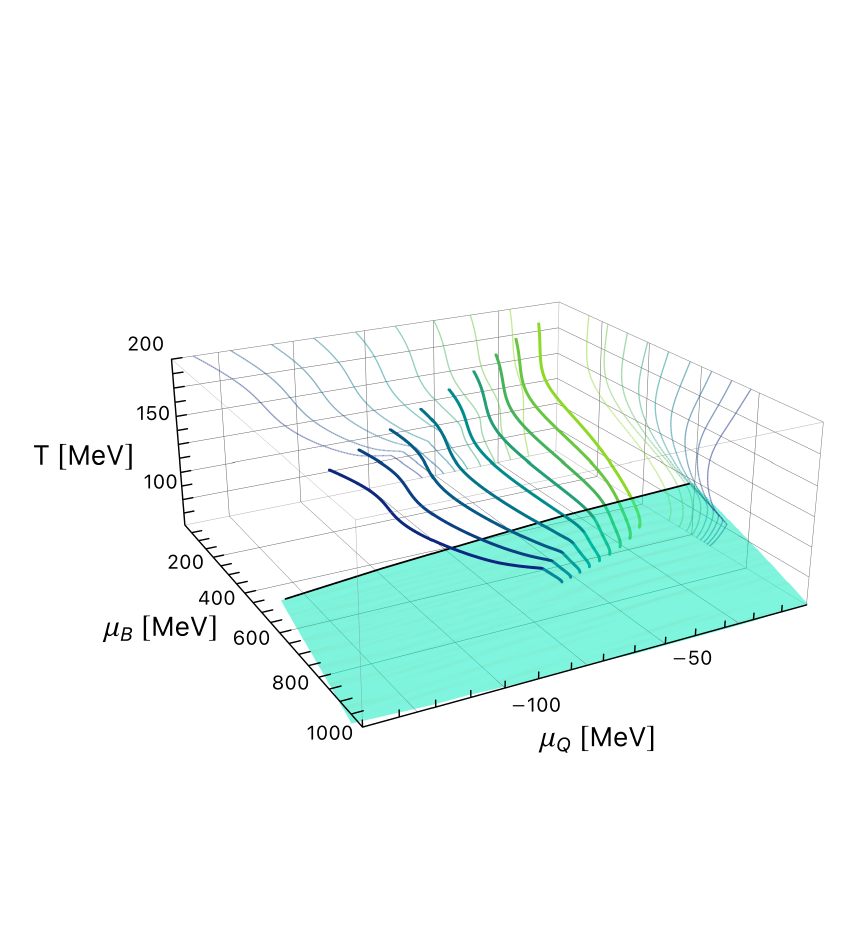}
   \caption{Cosmic trajectories that reach the first order phase transition surface. A lensing effect of the critical line is evident in the $\mu_B-T$ projection plane.}
    \label{lensing}
\end{figure}

This indicates that the contribution of the electric-charge chemical potential to the critical point (or critical line) does not significantly affect the nature of the cosmological QCD phase transition. In a logarithmic representation, the critical line is almost constant in $\mu_B$. Indeed, increasing $b$ makes it much easier to push the cosmic trajectories towards larger baryon chemical potential without requiring a correspondingly large $\mu_Q$ (even if $b\approx-l$ is considered), which would otherwise tend to drive the system out of the first-order region again.
As can be observed in \autoref{lensing}, once $b\sim 0.028$ the trajectory reaches the FOPT surface at $T\sim 114\,\mathrm{MeV}$, where the single-phase calculation is stopped since the description in terms of a single homogeneous phase no longer applies. All trajectories corresponding to larger values of $b$ therefore pass through the first-order region.

To reach a first-order phase transition we then need at least $b_{pt}=0.028$, which must be diluted to $b_{obs}=8.65\cdot 10^{-11}$ before the BBN, since, as discussed above, this is the earliest epoch at which we can measure it. We define the dilution factor $\mathcal{D}$ as \cite{Boeckel_2012}
\begin{equation}
   \mathcal{D}= \frac{b_i}{b_f}=\frac{0.028 }{8.65\cdot10^{-11}}\approx3.2\cdot 10^{8},\ 
\end{equation}
where $b_{i}$ is the baryon asymmetry needed to reach a FOPT, and $b_{f}$ is the observed (diluted) one.

To determine the required expansion, we calculate the number of $e$-folds necessary to achieve this dilution by reviewing the mechanism of little inflation. The total entropy $S$ and total net baryon number $N_B$ are conserved both before and after the transition. These quantities scale as $S \propto s a^3$ and $N_B \propto n_B a^3$, where $a$ is the scale factor. Consequently, the baryon asymmetry $b \equiv N_B / S = n_B / s$ remains constant during standard adiabatic expansion.

Before entering the supercooled regime, i.e., when the trajectory drops below the critical surface while remaining trapped in the metastable QGP false vacuum, the scale factor evolves as in a standard radiation-dominated universe, $a \propto 1/T$. Once the Universe becomes trapped in the false vacuum, this expansion becomes exponential, $a(t) \propto e^{H t}$, where $H$ is the Hubble parameter. At this stage, expansion accelerates, driving little inflation. When the bubble nucleation rate matches the expansion rate of the Universe ($\Gamma / H \approx 1$), the system transitions to the true minimum. The false vacuum energy is released into the plasma on a timescale much shorter than the expansion rate ($a \approx \text{const.}$), producing a rapid jump in both total entropy $S$ and temperature (reheating). The net comoving baryon number, $N_B$, is strictly conserved throughout the phase transition and subsequent reheating phase ($N_{Bi} = N_{Bf}$). Consequently, the massive injection of total entropy $S$ during reheating dilutes the initial baryon to entropy ratio, $b \equiv N_B / S$, to its actual observed value, $b_{\text{obs}}$. 

In the standard radiation-dominated epochs preceding and following the supercooled phase, the scale factor scales inversely with temperature ($a \propto 1/T$). By defining our initial and final states at the same reference temperature, the entropy densities before and after the transition are approximately equal ($s_i \approx s_f$). Using the definition $s \equiv S/a^3$, we can relate the jump in total entropy directly to the volumetric expansion of the Universe:
\begin{align}
    \frac{S_i}{S_f} &\approx \left(\frac{a_i}{a_f}\right)^3, \\
    \frac{b_f}{b_i} &= \frac{S_i}{S_f} \approx \left(\frac{a_i}{a_f}\right)^3 \approx 3.1 \times 10^{-9}.\ 
\end{align}
To achieve the required dilution, the scale factor must grow by a factor of $a_f / a_i \approx 687$, corresponding to a ratio of $a_i / a_f \approx 1.46 \times 10^{-3}$.

The first-order transition found for $b\gtrsim0.028$ does not by itself dilute the baryon asymmetry: within the present EoS the metastable plasma phase persists only a few MeV below the coexistence temperature, so the transition proceeds almost adiabatically and $b$ is conserved. Reconciling $b_{\rm QCD}\gtrsim0.028$ with $b_{\rm obs}$ therefore requires an entropy release of $\mathcal{D}=b_{\rm QCD}/b_{\rm obs}\approx3\cdot10^{8}$ from physics beyond the lattice-based EoS, such as the little-inflation scenario \cite{Boeckel_2012}, whose attainability within standard-model QCD has recently been questioned \cite{Jung:2026jlz}.
If such a phase exists, the corresponding expansion is $N=\ln\mathcal{D}/3\approx6.5$ $e$-folds. We do not model this phase here. 
Note that the same dilution affects the total lepton asymmetry and each $\ell_i$, leaving the ratios $\ell/b$ and $\ell_i/l$ unchanged.

One of the effects of the QCD critical point is the lensing effect of the hydrodynamic trajectories \cite{Dore_2022}. A very similar effect is also obtained here: in \autoref{lensing} and more evidently in \autoref{fig:bvaryngl}, especially in the $\mu_B-T$ plane, as long as the trajectories are in the crossover regime, the closer they get to the critical line, the more they group together and deform along its edge. A very similar effect has been observed with cosmic trajectories calculated with the functional QCD EoS \cite{Gao:2024fhm}.

\subsection{Pion condensation and gravitational wave imprints}
In the electro-phobic configuration ($\ell_e = 0$, $\ell_\mu = -\ell_\tau$), the electric charge chemical potential can reach large enough values to exceed the pion mass ($|\mu_Q| \geq m_\pi = 139.6\,\text{MeV}$), reaching the regime of pion Bose-Einstein condensation (BEC)\cite{Vovchenko_2021,Ferreira:2025zeu,ferreira2026signaturespioncondensationlepton}. As shown in Ref.~\cite{Vovchenko_2021}, a flavor-asymmetric configuration is necessary to cross this threshold within observational bounds: indeed, in \autoref{bounce} the threshold $|\mu_Q| = 139.6\,\text{MeV}$ (vertical dashed line) is crossed at high temperatures for $0.06 \leq \ell_\tau \leq 0.14$ (which then drop back below $m_\pi$ before $T = 80\,\text{MeV}$), whereas all trajectories with $\ell_\tau \geq 0.16$ remain at $|\mu_Q| > m_\pi$ throughout the entire temperature range down to $T = 80\,\text{MeV}$. Within our framework, the pion condensate enters explicitly through the QvdW-HRG matching condition at $T_{\text{low}} = 80\,\text{MeV}$ in Eq. \eqref{eq:pressure}, which directly defines the low-temperature thermodynamics for $\ell_\tau \geq 0.16$ and also influences the trajectory slopes at higher temperatures whenever $|\mu_Q| \geq m_\pi$. Conversely, for the symmetric configuration ($\ell_e = \ell_\mu = \ell_\tau = \ell/3$), $|\mu_Q|$ never reaches $m_\pi$, even in the non-standard scenario with $b = -\ell$ (\autoref{lensing}), where $|\mu_Q| \lesssim 120\,\text{MeV}$.

\begin{figure}[h]
    \centering
    {\includegraphics[width=0.7\linewidth]{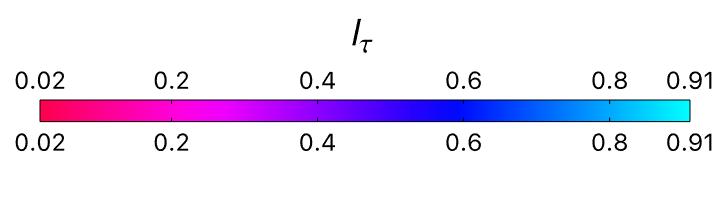}}
    \includegraphics[width=1.\columnwidth]{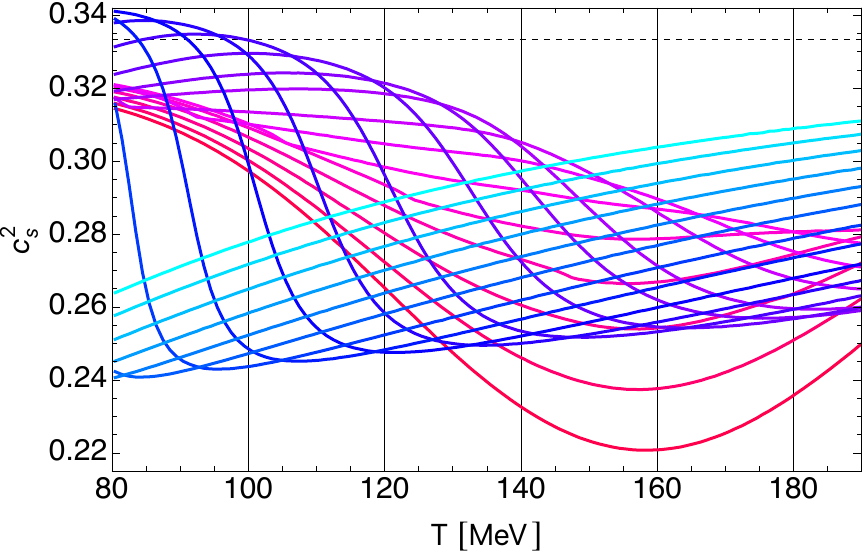}
    \caption{Squared speed of sound $c_s^2$ as a function of the temperature $T$ evaluated along the cosmic trajectories in the electro-phobic scenario ($\ell_e = 0$, $\ell_\mu = -\ell_\tau$, and $b = 8.65 \cdot 10^{-11}$) for the same values of $\ell_\tau \in [0.02, 0.91]$ and color coding shown in \autoref{bounce}. The horizontal dashed grey line indicates the conformal limit $c_s^2 = 1/3$.}
    \label{cs2}
\end{figure}

In \autoref{cs2} we show the squared speed of sound $c_s^2(T)$ along the electro-phobic trajectories of \autoref{bounce}. For $\ell_\tau = 0.02$ (lowest red curve), $|\mu_Q| < m_\pi$ and $c_s^2(T)$ exhibits the standard QCD crossover dip at $T \simeq 160\, \text{MeV}$ ($c_s^2 \approx 0.22$). For $\ell_\tau \geq 0.16$, where $|\mu_Q| > m_\pi$ down to $T = 80\,\text{MeV}$, increasing $\ell_\tau$ progressively shifts the minimum of $c_s^2(T)$ toward lower temperatures. For intermediate asymmetries ($0.28 \leq \ell_\tau \leq 0.46$), after passing this minimum, $c_s^2(T)$ undergoes a pronounced rise upon cooling toward $T = 80\,\text{MeV}$, where the pion-condensed HRG boundary condition applies, exceeding the conformal limit $c_s^2 = 1/3$ for $0.32 \leq \ell_\tau \leq 0.41$ and peaking at $c_s^2 \simeq 0.34$ right at the lower edge of the plot ($T = 80\,\text{MeV}$). Finally, for $\ell_\tau \geq 0.52$ (light-blue and cyan curves), the turnaround of $|\mu_Q|$ and the minimum of $c_s^2(T)$ shift down to $T \lesssim 80\,\text{MeV}$, so within the plotted window $c_s^2(T)$ decreases monotonically upon cooling.

A similar low-temperature stiffening is found in the equation-of-state parameter $\omega(T) = P(T)/\varepsilon(T)$ shown in \autoref{fig:PBH_l}(a), which also rises slightly above $1/3$ near $T = 80\,\text{MeV}$ for intermediate $\ell_\tau$, in agreement with the HRG plus pion condensation results of Ref. \cite{ferreira2026signaturespioncondensationlepton}. Although electro-phobic trajectories remain in the crossover regime and cannot generate gravitational waves via a FOPT, this stiffening can still leave an imprint in the nanohertz band probed by PTAs by modifying the low-frequency ``causality tail'' of gravitational waves produced prior to the QCD epoch \cite{ferreira2026signaturespioncondensationlepton}.

\label{GW}

\subsection{Primordial Black Holes formation over the QCD phase transition}
\label{sec:PBH}

\begin{figure*}[]
    \centering
    \begin{minipage}[b]{0.48\textwidth}
        \centering
        \includegraphics[width=\textwidth]{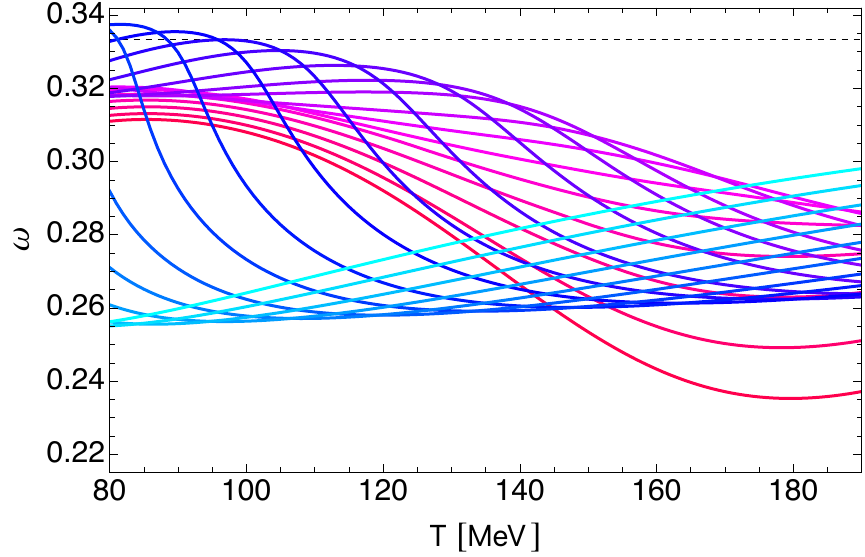} 
        \par\vspace{2pt}{\small (a) Equation of state $\omega=P/\rho$}
    \end{minipage}
    \hfill
    \begin{minipage}[b]{0.48\textwidth}
        \centering
        \includegraphics[width=\textwidth]{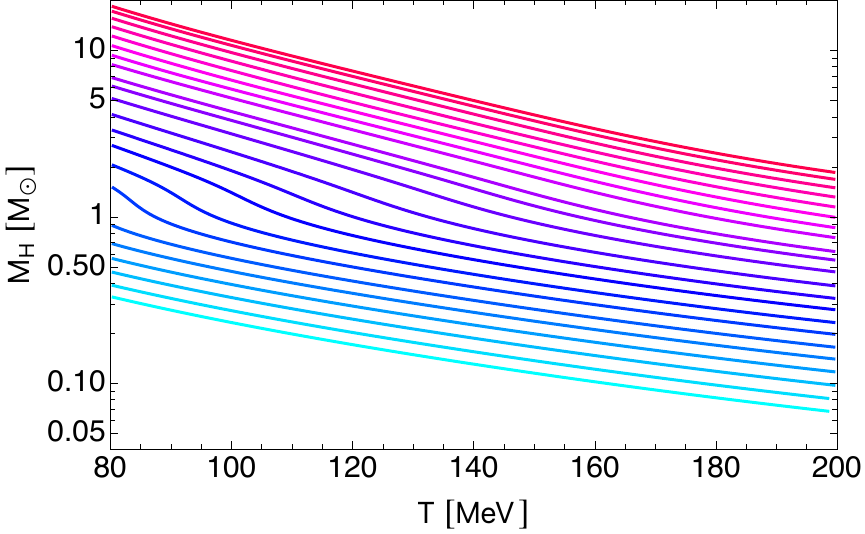}
        \par\vspace{2pt}{\small (b) Horizon Mass $M_H(T)$}
    \end{minipage}
    \vspace{0.4cm}
    \begin{minipage}[b]{0.48\textwidth}
        \centering
        \includegraphics[width=\textwidth]{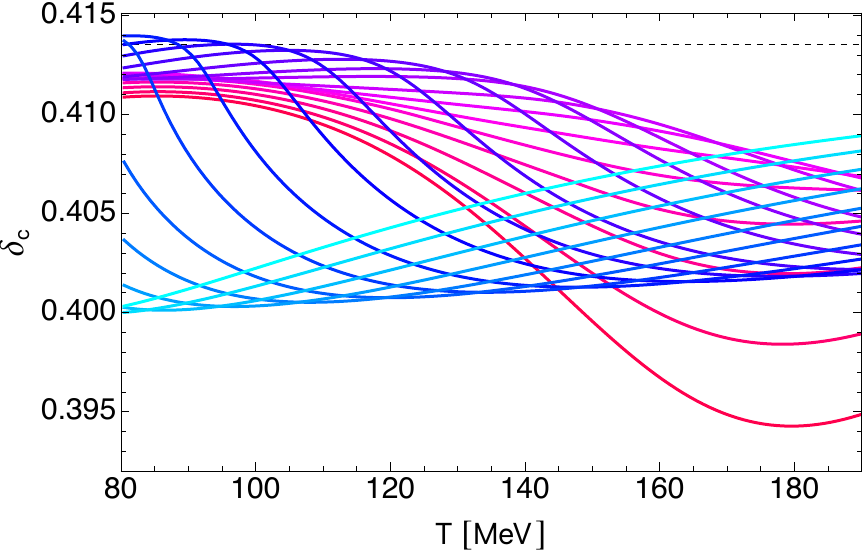} 
        \par\vspace{2pt}{\small (c) Collapse threshold $\delta_c(T)$}
    \end{minipage}
    \hfill
    \begin{minipage}[b]{0.48\textwidth}
        \centering
        \includegraphics[width=\textwidth]{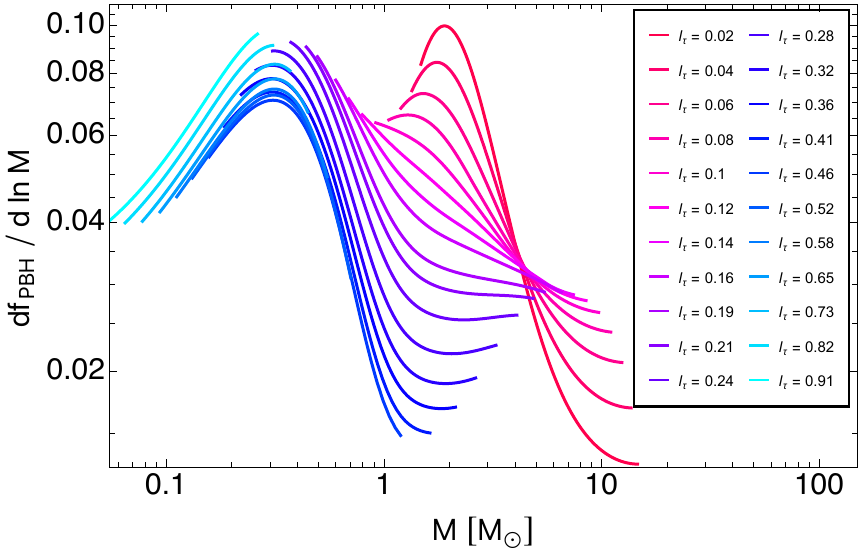}
        \par\vspace{2pt}{\small (d)Mass spectrum $df_{\mathrm{PBH}}/d\ln M$}
    \end{minipage}
  \caption{PBHs formation for the standard electro-phobic scenario ($b = 8.65 \cdot 10^{-11}$, $\ell_e = 0$, $\ell_\mu = -\ell_\tau$) with $\ell_\tau \in [0.02, 0.91]$. Horizontal dashed lines in panels (a) and (c) indicate the pure-radiation values $\omega = 1/3$ and $\delta_c(1/3) \simeq 0.4135$.}
    \label{fig:PBH_l}
\end{figure*}
The softening of the cosmic equation of state across the QCD epoch also leaves a direct imprint on the mass distribution of primordial black holes (PBHs) formed from the collapse of large horizon-size density perturbations \cite{Carr:1974nx, Carr:1975qj, Jedamzik:1996mr, Byrnes:2018clq}. Because gravitational collapse is opposed by radiation pressure, any reduction in the equation-of-state parameter $\omega(T) \equiv P(T)/\varepsilon(T)$, where $P(T)$ and $\varepsilon(T)$ are the total cosmic pressure and energy density along the trajectory, lowers the critical collapse threshold $\delta_c$ and exponentially enhances PBH formation \cite{Jedamzik:1996mr, Byrnes:2018clq, Carr:2019kxo}. From the Friedmann equation $H^2 = \rho/(3M_{\rm Pl}^2)$, with $M_{\rm Pl} = (8\pi G)^{-1/2} = 2.435 \times 10^{21}\,\mathrm{MeV}$, the PBH mass formed at horizon crossing is proportional to the Hubble horizon mass $M_H(T)$ \cite{Byrnes:2018clq, Gonin:2026nu}:
\begin{equation}
    M = \gamma M_H(T) = \gamma\,\frac{4\sqrt{3}\,\pi M_{\rm Pl}^3}{\sqrt{\rho(T)}},
    \label{eq:MH_exact}
\end{equation}
where we set $\gamma = 0.8$ \cite{Byrnes:2018clq} and evaluate the QCD energy density consistently via the Euler relation $\varepsilon_{\rm QCD} = T s_{\rm QCD} - P_{\rm QCD} + \mu_B n_B + \mu_Q n_Q$. Unlike the standard $\mu = 0$ approximation $M_H \propto g_*(T)^{-1/2} T^{-2}$\cite{Byrnes:2018clq, Bodeker:2020stj}, Eq.~\eqref{eq:MH_exact} remains exact at arbitrary chemical potentials. We compute the collapse threshold in the comoving gauge using the analytical prescription of Ref. \cite{Harada:2013epa},
\begin{equation}
    \delta_c(\omega) = \frac{3(1 + \omega)}{5 + 3\omega}\sin^2\!\left(\frac{\pi\sqrt{\omega}}{1 + 3\omega}\right),
    \label{eq:delta_c_HYK}
\end{equation}
which gives $\delta_c(1/3) \simeq 0.4135$ for pure radiation; relativistic hydrodynamical simulations across the QCD transition confirm that collapsing perturbations respond to the Hubble-averaged pressure, yielding a smooth threshold profile well approximated by $\delta_c(\omega(T))$ \cite{Musco:2012au, Byrnes:2018clq, Franciolini:2022tfm, Escriva:2022bwe, Musco:2023dak}. Within the Press--Schechter formalism \cite{Press:1973iz, Carr:1975qj} for Gaussian perturbations with root-mean-square amplitude $\sigma_{\rm rms}(M) = A(M/M_\odot)^{(1-n_s)/4}$ and $n_s = 0.9649$\cite{Planck:2018vyg},
the present-day differential fraction of cold dark matter (CDM) in PBHs is \cite{Nakama:2016gzw, Byrnes:2018clq, Bodeker:2020stj, Gonin:2025qcd}
\begin{align}
    \frac{df_{\rm PBH}(M)}{d\ln M} &\simeq 2.4\,\beta(M)\sqrt{\frac{M_{\rm eq}}{M}} \\
    \beta(M) &= \operatorname{erfc}\!\left[\frac{\delta_c(\omega(T(M)))}{\sqrt{2}\,\sigma_{\rm rms}(M)}\right],
    \label{eq:dfPBH}
\end{align}
where $\beta(M)$ is the collapse fraction at formation, $M_{\rm eq} \simeq 2.8 \times 10^{17}\,M_\odot$ is the horizon mass at matter--radiation equality, and $2.4 \simeq 2(1 + \Omega_b/\Omega_{\rm CDM})$. Following Refs. \cite{Carr:2019kxo, Gonin:2025qcd, Gonin:2026beyond, Gonin:2026nu}, on each trajectory we calibrate the amplitude $A = A_{\rm opt}$ so that the integrated PBH fraction over $80\,\mathrm{MeV} \le T \le 200\,\mathrm{MeV}$ equals $f_{\rm PBH} = \int (df_{\rm PBH}/d\ln M)\,d\ln M = 0.1$ (it constitutes $10\%$ of DM).

\begin{figure*}[]
    \centering
    \begin{minipage}[b]{0.48\textwidth}
        \centering
        \includegraphics[width=\textwidth]{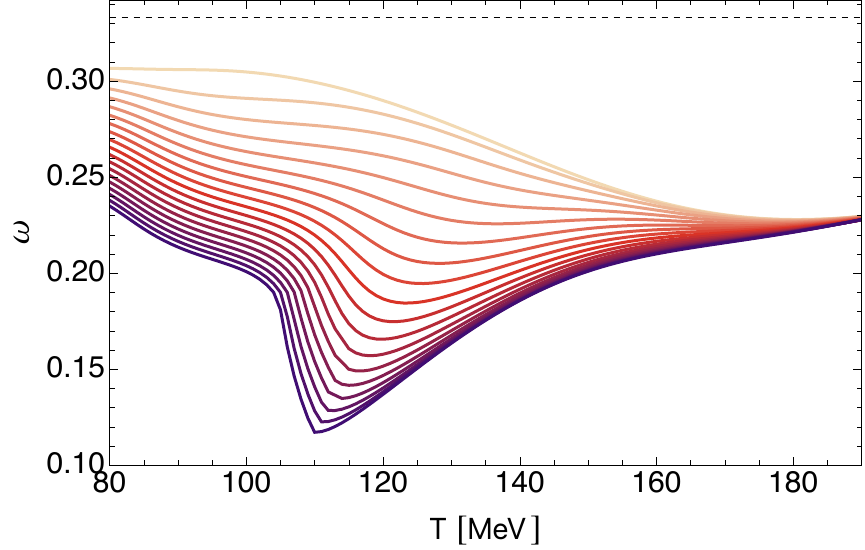} 
        \par\vspace{2pt}{\small (a) Equation of state $\omega=P/\rho$}
    \end{minipage}
    \hfill
    \begin{minipage}[b]{0.48\textwidth}
        \centering
        \includegraphics[width=\textwidth]{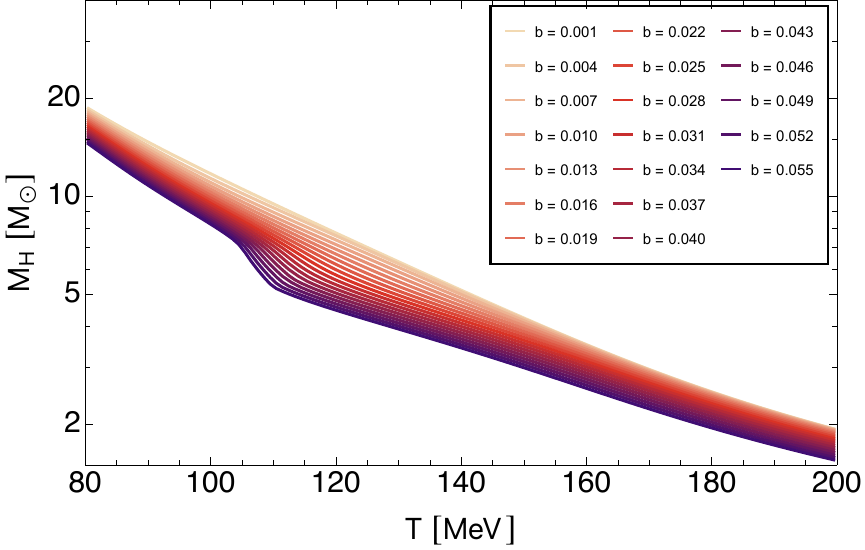}
        \par\vspace{2pt}{\small (b) Horizon Mass $M_H(T)$}
    \end{minipage}
    \vspace{0.4cm}
    \begin{minipage}[b]{0.48\textwidth}
        \centering
        \includegraphics[width=\textwidth]{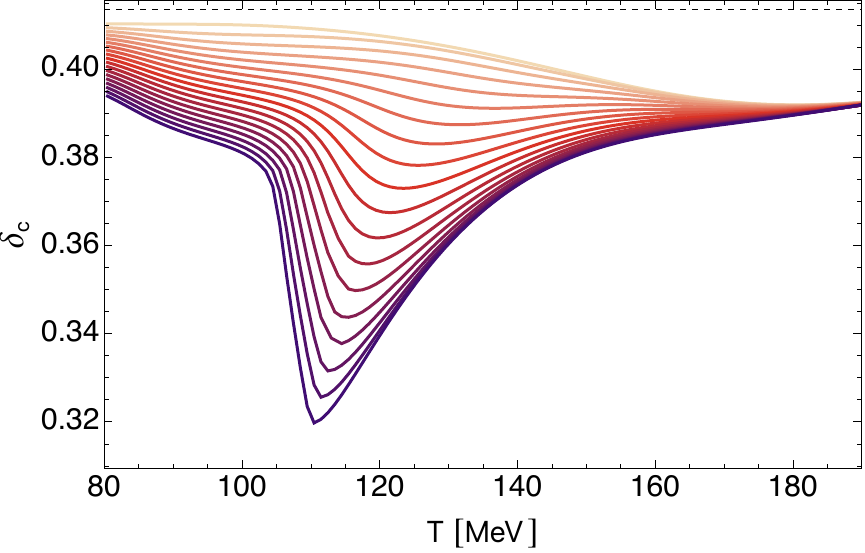} 
        \par\vspace{2pt}{\small (c) Collapse threshold $\delta_c(T)$}
    \end{minipage}
    \hfill
    \begin{minipage}[b]{0.48\textwidth}
        \centering
        \includegraphics[width=\textwidth]{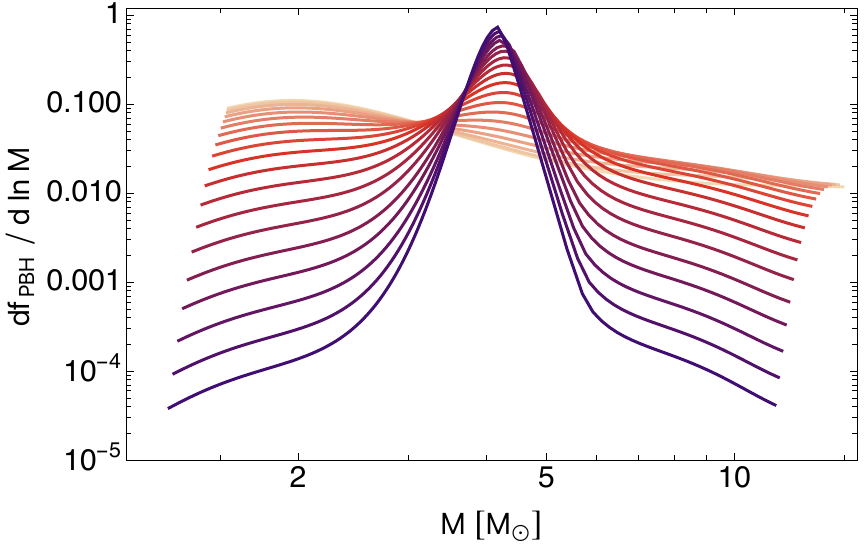}
        \par\vspace{2pt}{\small (d)Mass spectrum $df_{\mathrm{PBH}}/d\ln M$}
    \end{minipage}
    \caption{PBHs formation for the non-standard scenario with fixed symmetric lepton asymmetry $\ell = -10 \cdot 10^{-10}$ and varying baryon asymmetry $b \in [0.001, 0.055]$.}
    \label{fig:PBH_b}
\end{figure*}

In the standard electro-phobic scenario ($b = 8.65 \times 10^{-11}$, $\ell_e = 0$, $\ell_\mu = -\ell_\tau$), shown in \autoref{fig:PBH_l},
small asymmetries ($\ell_\tau = 0.02$) reproduce the standard $\mu \approx 0$ QCD crossover softening ($\omega_{\rm min} \simeq 0.233$, $\delta_{c,{\rm min}} \simeq 0.394$ at $T \simeq 180\,\mathrm{MeV}$), generating the familiar peak at $M \simeq 1.9\,M_\odot$ ($A_{\rm opt} \simeq 0.0607$). Increasing $\ell_\tau$ up to $0.91$ extends the analysis deep into the pion-condensed regime ($|\mu_Q| \ge m_\pi$ for $\ell_\tau \ge 0.06$ \cite{Vovchenko_2021, diclemente2025upperboundcosmicbaryon}),
beyond the domain accessible to previous Taylor-expanded or quasiparticle studies \cite{Bodeker:2020stj, Gonin:2025qcd, Gonin:2026beyond}. The extra energy density of degenerate leptons and condensed pions substantially increases $\varepsilon(T)$ at fixed $T$, shrinking $M_H(T)$ by over an order of magnitude (\autoref{fig:PBH_l}-b) and shifting the PBH peak into the sub-solar window $M \simeq 0.3\text{--}0.8\,M_\odot$ (\autoref{fig:PBH_l}-d)), a prime target for microlensing surveys and LIGO--Virgo--KAGRA (LVK) sub-solar searches \cite{EROS-2:2006ryy, Mroz:2024mse, LIGOScientific:2022hai, Carr:2026rev}. Concurrently, the leptonic and pion-BEC stiffening softens the dip in $\omega(T)$ ($A_{\rm opt} \in [0.0607, 0.0619]$) and pushes $\omega(T) \gtrsim 1/3$ near $T \simeq 80\,\mathrm{MeV}$ for intermediate $\ell_\tau$, raising $\delta_c$ above $\delta_c(1/3)$ (\autoref{fig:PBH_l}-c) and suppressing the high-mass tail.
\begin{figure*}[]
    \centering
    \begin{minipage}[b]{0.48\textwidth}
        \centering
        \includegraphics[width=\textwidth]{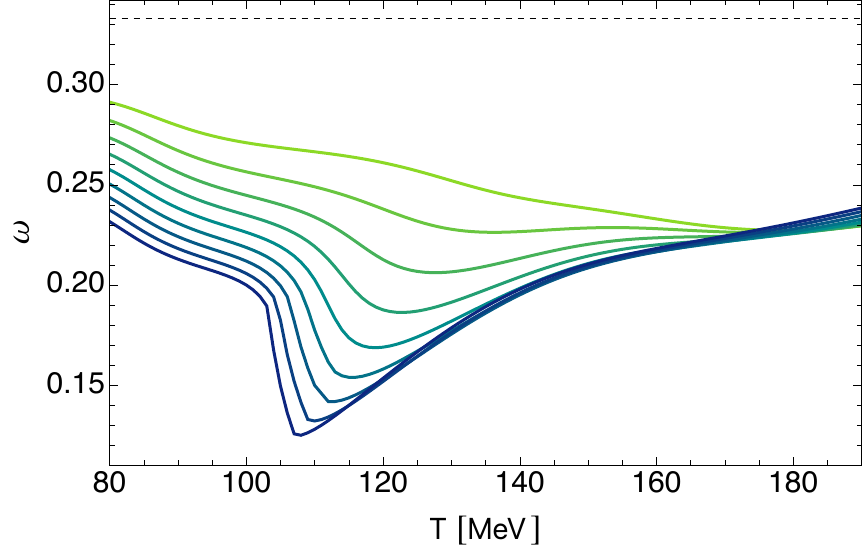} 
        \par\vspace{2pt}{\small (a) Equation of state $\omega=P/\rho$}
    \end{minipage}
    \hfill
    \begin{minipage}[b]{0.48\textwidth}
        \centering
        \includegraphics[width=\textwidth]{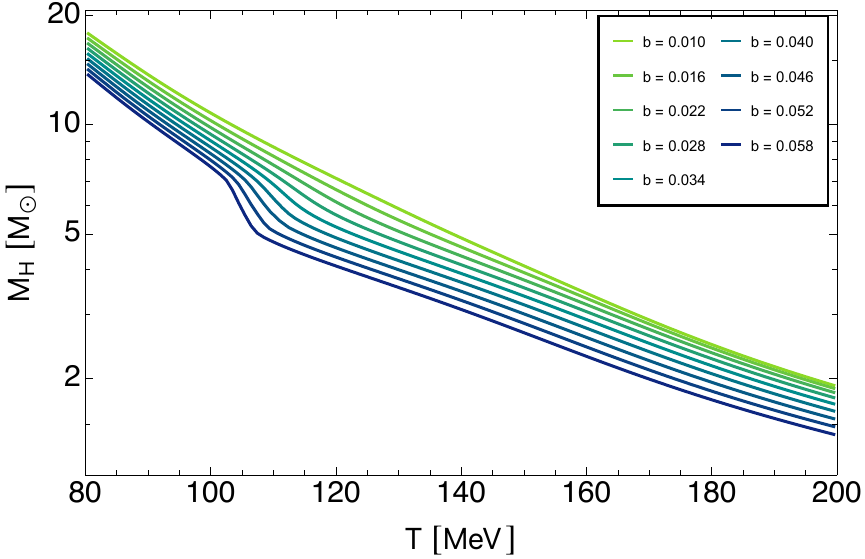}
        \par\vspace{2pt}{\small (b) Horizon Mass $M_H(T)$}
    \end{minipage}
    \vspace{0.4cm}
    \begin{minipage}[b]{0.48\textwidth}
        \centering
        \includegraphics[width=\textwidth]{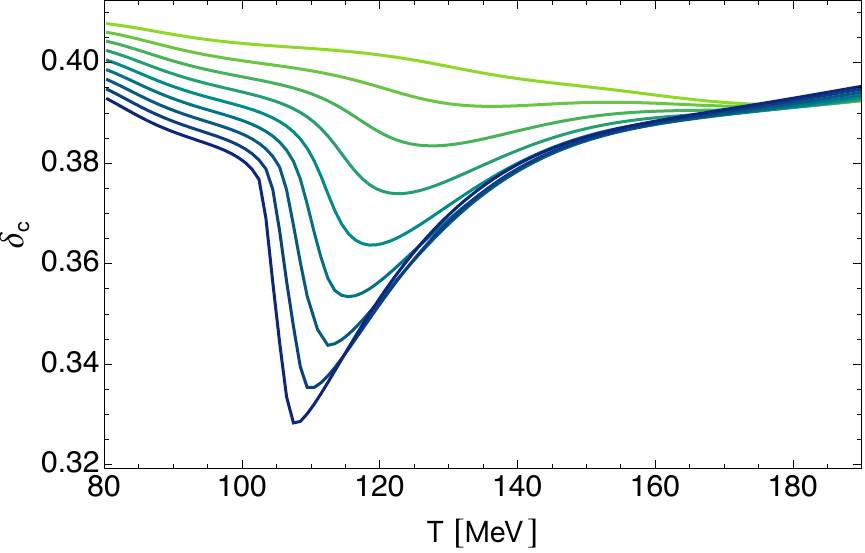} 
        \par\vspace{2pt}{\small (c) Collapse threshold $\delta_c(T)$}
    \end{minipage}
    \hfill
    \begin{minipage}[b]{0.48\textwidth}
        \centering
        \includegraphics[width=\textwidth]{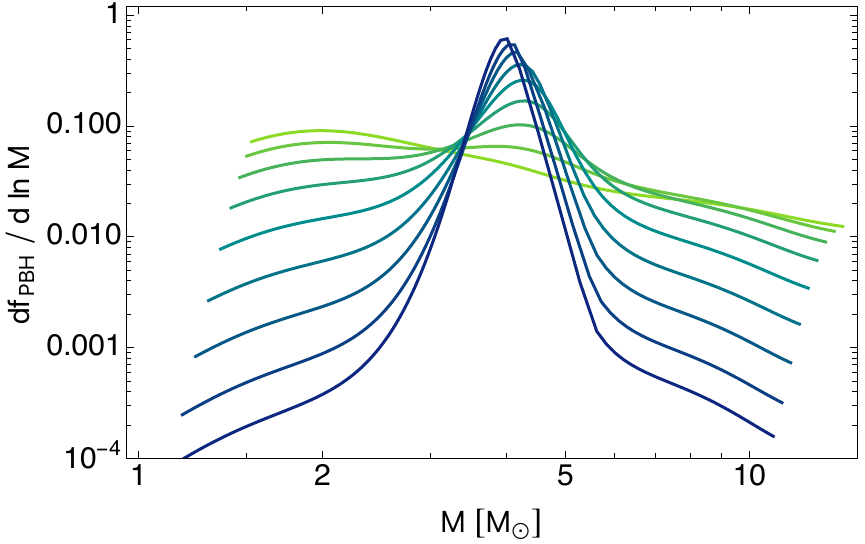}
        \par\vspace{2pt}{\small (d)Mass spectrum $df_{\mathrm{PBH}}/d\ln M$}
    \end{minipage}
    \caption{PBHs formation for the non-standard scenario with symmetric lepton asymmetries ($\ell_e = \ell_\mu = \ell_\tau = \ell/3$) and $b = -\ell \in [0.010, 0.058]$.}
    \label{fig:PBH_bl}
\end{figure*}
Conversely, in the non-standard scenarios with large baryon asymmetry, both for $b = -\ell$ (\autoref{fig:PBH_bl}) and for fixed $\ell = -10 \times 10^{-10}$ (\autoref{fig:PBH_b}), where standard Taylor expansions break down ($\mu_B/T \gtrsim 0.1$) \cite{Formaggio:2025nde, Gonin:2026beyond}, critical lensing and the crossing of the FOPT surface ($b \geq 0.028$) at $T \simeq 113\text{--}114\,\mathrm{MeV}$ cause a pronounced softening down to $\omega_{\rm min} \simeq 0.11\text{--}0.12$ and $\delta_{c,{\rm min}} \simeq 0.32$ (\autoref{fig:PBH_bl}-a,c and \autoref{fig:PBH_b}-a,c). Because this transition occurs at lower temperature where $M_H \simeq 5.5\text{--}6\,M_\odot$, the broad $\sim 2\,M_\odot$ crossover peak turns into a sharp, almost monochromatic spike at $M \simeq 4.0\text{--}4.3\,M_\odot$(\autoref{fig:PBH_bl}-d and \autoref{fig:PBH_b}-d), near the upper edge of the LVK lower mass gap \cite{Bagui:2023zrj, LIGOScientific:2025gwtc4}, while $A_{\rm opt}$ drops from $0.0605$ to $0.0519$ regardless of $\ell$ (corresponding to an abundance enhancement of $\sim 10^4$ at fixed $A$). Note that Eq. \eqref{eq:dfPBH} applies directly without post-transition dilution if these high-$b$ trajectories are realized locally within rare superhorizon patches, such as Affleck--Dine high-baryon bubbles \cite{Dolgov:1992pu, Kasai:2022vhq}, in a background Universe with $b_{\rm obs}$. In a globally high-$b$ scenario diluted after the transition by $\mathcal{D} \approx 3.9 \times 10^8$ \cite{Boeckel_2012}
(Sec. \ref{littleinflation}), the pre-existing PBH density is diluted by the same factor $\mathcal{D}$, so the narrow $\sim 4.2\,M_\odot$ spectral shape is preserved with $A$ effectively rescaled (potentially alongside non-adiabatic FOPT production channels for $b \ge 0.028$~\cite{Baker:2021nyl, Lewicki:2023ioy}).

\section{Conclusions}
\label{conclusions}

In this work, we have investigated the cosmological QCD phase transition by calculating, for the first time, cosmic trajectories within the QCD phase diagram using an equation of state that predicts a first-order phase transition with a critical surface in a four-dimensional parameter space ($T-\mu_B-\mu_Q-\mu_S$), specifically the constant-entropy contour-based EoS~(CEC EoS), which for $\mu_S = 0$ ($\beta$-equilibrium) defines a critical line and a two-dimensional first-order transition surface in $(T, \mu_B, \mu_Q)$.
This framework allowed us to account for the $\mu_Q$-dependence of the QCD critical line and determine how it affects the nature of the cosmological phase transition along the cosmic trajectories.
In particular, we addressed whether a sufficiently high electric charge chemical potential can drive the system through a first-order transition, even when the baryon chemical potential remains constrained.

We investigated three main scenarios: (i) the minimal standard scenario, 
using the observed small baryon asymmetry $b=8.65\cdot 10^{-11}$ and sphaleron-based lepton asymmetry $\ell=\ell_{msc}=-51/28 b$; (ii) the standard scenario, where the standard value of baryon asymmetry is preserved but the relationship between $b$ and $\ell$ is relaxed, with the three lepton asymmetries bounded only by observational constraints; 
and (iii) the non-standard scenario, where both $b$ and $\ell_i$ are assumed to be free from any observational or theoretical constraints. 

Previous studies\cite{diclemente2025upperboundcosmicbaryon,Formaggio:2025nde,wygas2018cosmic,Oldengott_2017,PhysRevD.105.123533}, employing a second order Taylor expanded lattice QCD EoS with $2+1$ flavors, recently extended to fourth order with $2+1+1$ flavors, have shown that lepton asymmetries can push the trajectories up to $\mu_B\sim 150\, \mathrm{MeV}$ without ever inducing a first-order phase transition. We found that in both the minimal standard and standard scenarios, assuming a symmetric configuration of the lepton asymmetries, the trajectories lie at an energy scale such that the critical line along the $\mu_B$ direction remains constant with respect to $\mu_Q$. This yields no significant deviations from previous findings.

Conversely, in the asymmetric configuration, where the relevant energy scale for $\mu_{Q,B}$ reaches hundreds of MeV, the CEC EoS predicts a distinct variation of the critical line across the $\mu_B-\mu_Q$ plane. Nevertheless, the trajectories remain entirely within the crossover regime, corroborating earlier results in the literature. Consequently, we conclude that lepton asymmetries alone cannot induce a cosmological QCD FOPT.

To achieve a FOPT, it is necessary to relax the constraint on the baryon asymmetry $b$. Indeed, we demonstrate that for $b \geq 0.028$, the trajectories intersect the first-order phase transition surface regardless of the specific lepton asymmetries, which in this regime play only a minor, fine-tuning role. To account for such an elevated baryon asymmetry, one must introduce a dilution mechanism that reduces $b$ from its initial state to the currently observed value, bridging a gap of several orders of magnitude. Within the present EoS the transition itself does not provide this dilution, because its metastable phase persists only a few MeV below coexistence. If the entropy is released by little inflation \cite{Boeckel_2012}, an initial baryon asymmetry of $b=0.028$ requires approximately $N \approx 6.5$ $e$-folds of expansion.

We also scanned a variety of parameter configurations to determine the main effects of each asymmetry on the trajectories. We show that the primary effect of lepton asymmetries is driven by the electric charge chemical potential. As a secondary effect, this leads to an increase in the baryon chemical potential, because more baryons are required to balance the electric charge introduced in the lepton sector. However, this mechanism does not effectively push $\mu_B$ to high values. 
Conversely, increasing only the baryon asymmetry $b$ makes it much easier to reach high $\mu_B$. This produces a similar charge-balancing effect on $\mu_Q$, as more leptons are needed to offset the electric charge introduced by the baryons. However, because the quarks carry fractional electric charges, fewer leptons are required to balance the baryonic charge, resulting in only a minimal variation in $\mu_Q$ when $b$ is increased. For the opposite reason, if the lepton density increases due to $\ell_\alpha$, more baryons are required to balance the electric charge introduced by the leptons. Consequently, lepton asymmetries show a stronger secondary effect on $\mu_B$ than $b$ has on $\mu_Q$.

We observe that, the closer the trajectories are to the critical line, the more they are attracted by it, resulting in a ``lensing effect'' similar to the one predicted for isentropic trajectories in heavy-ion collisions \cite{Dore_2022}.

Furthermore, in the electro-phobic scenario for $\ell_\tau \geq 0.16$, $|\mu_Q|$ remains above the pion mass down to $T = 80\,\mathrm{MeV}$, where the pion-condensed HRG matching condition applies and stiffens the equation of state upon cooling~\cite{Vovchenko_2021, ferreira2026signaturespioncondensationlepton}. For intermediate asymmetries ($0.32 \leq \ell_\tau \leq 0.41$), this stiffening drives $c_s^2$ and $\omega = P/\varepsilon$ above the conformal limit $1/3$ at $T \lesssim 100\,\mathrm{MeV}$, peaking at $c_s^2 \simeq 0.34$ at $T = 80\,\mathrm{MeV}$. Even without a QCD FOPT, this effect can leave an imprint on the low-frequency causality tail of the nanohertz gravitational-wave background probed by PTAs~\cite{ferreira2026signaturespioncondensationlepton}.

Finally, we explored the implications of the CEC equation of state for the production and mass spectrum of primordial black holes (PBHs) across the QCD transition~\cite{Jedamzik:1996mr, Byrnes:2018clq, Bodeker:2020stj, Musco:2023dak, Gonin:2026beyond}. In the standard electro-phobic scenario, increasing $\ell_\tau$ drives the system into the pion-condensed regime~\cite{Vovchenko_2021, Ferreira:2025zeu}, where the additional energy density of degenerate leptons and condensed pions shrinks the Hubble horizon and shifts the canonical solar-mass QCD peak ($\sim 1.9\,M_\odot$)~\cite{Jedamzik:1996mr, Byrnes:2018clq} down into the sub-solar window ($M \simeq 0.3\text{--}0.8\,M_\odot$), while low-temperature stiffening suppresses the high-mass tail. Conversely, in the non-standard scenario with large baryon asymmetry ($b \geq 0.028$), critical lensing and the crossing of the first-order transition surface at $T \simeq 113\text{--}114\,\mathrm{MeV}$ induce a strong thermodynamic softening ($\omega_{\rm min} \simeq 0.11\text{--}0.12$), transforming the broad crossover distribution into a sharp, nearly monochromatic spike at $M \simeq 4.0\text{--}4.3\,M_\odot$ regardless of the lepton asymmetry. Together with gravitational-wave observations~\cite{Afzal_2023, Bagui:2023zrj}, sub-solar and stellar-mass PBH searches~\cite{EROS-2:2006ryy, LIGOScientific:2022hai, Mroz:2024mse, LIGOScientific:2025gwtc4, Carr:2026rev} thus provide complementary observational probes of cosmological asymmetries during the QCD epoch.
\begin{acknowledgements}
This material is based upon work supported by the National Science Foundation under grants No. PHY-2208724, PHY-2116686, PHY-2514763, PHY-2621752 and PHY-2623480, and within the framework of the MUSES collaboration, under Grant No. OAC-2103680. This material is also based upon work supported by the U.S. Department of Energy, Office of Science, Office of Nuclear Physics, under Awards Number DE-SC0022023 and DE-SC0026065, as well as by the National Aeronautics and Space Administration (NASA) under Award Number 80NSSC24K0767. HS
is also supported by the Deutsche Forschungsgemeinschaft
(DFG, German Research Foundation) through the CRC-
TR 211 “Strong-interaction matter under extreme condi-
tions” Project No. 315477589–TRR 211.  
\end{acknowledgements}

\bibliographystyle{apsrev4-2}
\bibliography{main}

\end{document}